\documentclass{aa}  
\usepackage{graphicx}
\usepackage{txfonts}

\usepackage{xcolor}

\usepackage{natbib,twoopt}
\usepackage[breaklinks=true]{hyperref} 
\bibpunct{(}{)}{;}{a}{}{,} 
\makeatletter
\newcommandtwoopt{\citeads}[3][][]{\href{http://adsabs.harvard.edu/abs/#3}%
{\def\hyper@linkstart##1##2{}%
\let\hyper@linkend\@empty\citealp[#1][#2]{#3}}}
\newcommandtwoopt{\citepads}[3][][]{\href{http://adsabs.harvard.edu/abs/#3}%
{\def\hyper@linkstart##1##2{}%
\let\hyper@linkend\@empty\citep[#1][#2]{#3}}}
\newcommandtwoopt{\citetads}[3][][]{\href{http://adsabs.harvard.edu/abs/#3}%
{\def\hyper@linkstart##1##2{}%
\let\hyper@linkend\@empty\citet[#1][#2]{#3}}}
\newcommandtwoopt{\citeyearads}[3][][]%
{\href{http://adsabs.harvard.edu/abs/#3}
{\def\hyper@linkstart##1##2{}%
\let\hyper@linkend\@empty\citeyear[#1][#2]{#3}}}
\makeatother

\begin{document} 
\title{Gaia colour-magnitude diagrams of young open clusters\thanks{Partially based on observations made with the Nordic Optical Telescope under Fast-Track program 61-406.}}
\subtitle{Identification in the UBC catalogue and a comparison of manual and automated analysis} 

\author{I.~Negueruela
       \inst{1,2}
       \and
       A.~de Burgos\inst{3,4}
       }
\institute{Departamento de F\'{\i}sica Aplicada, Facultad de Ciencias, Universidad de Alicante,\\ Carretera de San Vicente s/n, E03690, San Vicente del Raspeig, Spain\\
\email{ignacio.negueruela@ua.es}
\and
Instituto Universitario de In vestigaci\'on Inform\'atica, Universidad de Alicante
\and
Universidad de La Laguna, Dpto. Astrof\'{\i}sica, E-38206 La Laguna, Tenerife, Spain
\and
Instituto de Astrof\'{\i}sica de Canarias, E-38205 La Laguna, Tenerife, Spain\\
\email{abel.burgos@iac.es}
}
\date{Received; accepted }
%
\abstract
 {Automated analysis of \textit{Gaia} astrometric data has led to the discovery of many new high-quality open cluster candidates. With a good determination of their parameters, these objects become excellent tools to investigate the properties of our Galaxy.}  
 {We explore whether young open clusters can be readily identified from \textit{Gaia} data alone by studying the properties of their \textit{Gaia} colour-magnitude diagrams. We also want to compare the results of a traditional cluster analysis with those of automated methods.}
 {We selected three young open cluster candidates from the UBC catalogue, ranging from a well-populated object with a well-defined sequence to a poorly-populated, poorly-defined candidate. We obtained classification spectra for the brightest stars in each. We redetermined members based on EDR3 data and fitted isochrones to derive age, distance and reddening.}
 {All three candidates are real clusters with age below 100~Ma. UBC~103 is a moderately populous cluster, with an age around 70~Ma. At a distance of $\sim 3\:$kpc, it forms a binary cluster with the nearby NGC~6683. UBC~114 is a relatively nearby ($\sim1.5\:$kpc) poorly-populated cluster containing two early-B stars. UBC~587 is a dispersed, very young ($\leq 10\:$Ma) cluster located at $\sim3\:$kpc, behind the Cygnus~X region, and may be a valuable tracer of the Orion arm.}
 {The {\tt OCfinder} methodology for the identification of new open clusters is extremely successful, with even poor candidates resulting in interesting detections. The presence of an almost vertical photometric sequence in the \textit{Gaia} colour-magnitude diagram is a safe way to identify young open clusters. Automated methods for the determination of cluster properties give approximate solutions, but are still subject to some difficulties. There is some evidence suggesting that artificial intelligence systems may systematically underestimate extinction, which may impact in the age determination.}
\maketitle
%


\section{Introduction}

Open clusters are valued natural laboratories, allowing the study of stellar evolution, Galactic structure and chemical gradients, among other current topics. The advent of \textit{Gaia} astrometric and photometric data \citep{brown18} resulted in substantial advance in their knowledge. Current astrometric data allow a much better definition of membership than any previous datasets, leading to stronger constraints on stellar evolution, while future releases will in all likelihood improve our understanding of dynamical processes in these stellar systems.

In addition, the wealth of high-quality data has resulted in the discovery of many new cluster candidates, which, if confirmed, could lead to the doubling of the number of clusters known. Most of these cluster candidates have been found by means of automated methods, which use a variety of strategies, from visual searches \citep[e.g.][]{ferreira20} and supervised exploration of regions where a result is anticipated \citep[e.g.][]{maiz_villa1}  to friends-of-friends algorithms \citep[e.g.][]{liupang19} and machine-learning Bayesan algorithms \citep[e.g.][]{hao20,he20} over large regions of the sky.
Until now, by far the most successful exploration is that carried out at the University of Barcelona by using the {\tt OCfinder} method \citep{castro-ginard19}. Based on DR2 data,  \citet{castro-ginard20} 
presented almost 600 high quality new\footnote{Although a few of them can be identified with previously named, but unstudied, clusters, the vast majority represent new detections.} open cluster candidates, representing an increase by more than one third in the number of open clusters known. An even higher number have recently been discovered in EDR3 data \citep{castro-ginard22}, while other authors have reported similar numbers of new discoveries \citep[e.g.][]{hao22,he22}. These new cluster candidates discovered in \textit{Gaia} data now rival in number the clusters known before its launch. In most cases, however, the candidates have not been validated with other techniques. Here we will concentrate on the candidates found by the {\tt OCfinder} method. As these are named UBC (University of Barcelona Cluster) followed by a sequential number, we will collectively refer to them as UBC clusters.

Beyond the value of discovery, the new clusters may be used to study the structure, kinematics  and/or chemical  properties  of the  Milky  Way, depending on their age and the number and nature of their members. Thus, the new large catalogues of cluster candidates only become truly valuable when the basic parameters of the clusters are estimated. To handle such large numbers of objects, methods based on machine-learning algorithms are almost mandatory. As an example, \citet{cantat20} estimated parameters in a homogenous way for a very large sample of Milky Way open clusters, which included a sizable fraction of the DR2 UBC clusters \citep{castro-ginard20}. Their method is based on comparison to clusters of known parameters, and in most cases results in a fair agreement with previous determinations. Its accuracy and the possibility of systematic effects have not been assessed yet, though.

Independently of their age and nature, the reasons why these new clusters had not been found before are, in most cases, easy to identify. Some lie in areas of very high stellar density (cf. \citealt{ferreira20}; or see the case of Valparaiso~1 = UBC~106, independently -- and serendipitously -- discovered by \citealt{negueruela21}). Although \citet{castro-ginard22} conclude that their method preferentially identifies objects of limited size in the sky, most are dispersed groups, with no obvious central concentration that might strike the eye. Among these, some are low-density groups at moderately large distances, which may more readily be identified as young associations than open clusters (for instance, UBC~609, which includes the previously known compact clusters Waterloo~1 and Meyer~2, as well as many stars in the surrounding area). 

\begin{figure}
 \centering
 \resizebox{\columnwidth}{!}{\includegraphics[clip]{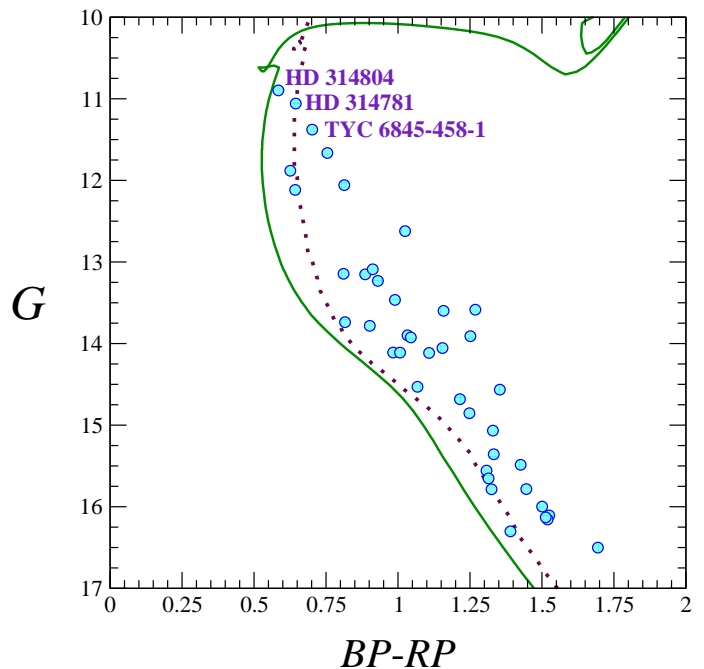}}
 \caption{\textit{Gaia} DR2 CMD for high-probability members of a ``typical'' UBC cluster, UBC~94. For reference, catalogued stars are named in the diagram. The green solid line represents the preferred isochrone fit according to \citet[280~Ma, $A_V=1.32$]{cantat20}. The purple dotted line is an alternative fit with a 150~Ma, $A_{V}=1.7$ isochrone. The lack of evolved stars allows for very different solutions.
              \label{typical}}%
    \end{figure}

Many other  UBC clusters are relative nearby, small groups of intermediate age, likely remnants of clusters that have already lost all their massive stars and are in the process of dissolution. A typical example of this kind of candidate may be UBC~94, whose \textit{Gaia} colour-magnitude diagram (CMD) is shown in Fig.~\ref{typical} for illustration. With their machine-learning approach, \citet{cantat20} derive a distance modulus $DM = 10.5$~mag (driven by the \textit{Gaia} DR2 parallax) with $A_{V}=1.32$, and an age of 280~Ma, but it is difficult to evaluate the goodness of such a fit, because the CMD is completely featureless. There is not a single evolved star and no guide to the position of the turn-off. As an example, we include a 150~Ma isochrone, with $A_{V}= 1.7$, which provides an equally valid (if not better) "fit" to the data for the same distance. Indeed, younger ages could also be accommodated simply by assuming that the upper main sequence is not populated, given the low number of members. The \textit{Gaia} CMD on its own does not contain enough information to determine an age. In this particular case, the three brightest stars are sufficiently bright to be included in catalogues. Two of them have HD classifications as A0 and A2, which supports the older age, even if we assume that they are giants. But such information is not available in most other cases, leading to an indetermination in the cluster properties. An artificial-intelligence algorithm, on the other hand, will always come back with an age for the cluster.

As discussed by \citet{cantat20}, the presence of red clump stars breaks the degeneracy between age and reddening and permits a quick and easy way to correctly assign ages. When no such stars are present, age determination is more difficult. Unfortunately, this is generally the case not only for poorly populated clusters, but also for young clusters, which have particularly high value, as they can be used to trace recent star formation, and provide information on Galactic structure, testing different models for the formation of spiral arms \citep[e.g.][]{castro-ginard21}. Moreover, the most luminous stars in young clusters will be OB stars, which have a high value in themselves, as probes of stellar evolution. In this paper, we explore the possibility that young open clusters can be identified among the new candidates by the shape of their \textit{Gaia} CMDs alone. In Section~\ref{sec:clusters}, we discuss the features that allow the identification of young open clusters in a \textit{Gaia} CMD. In Sect.~\ref{section.obs_astrometry}, we describe how we selected three young cluster candidates from the UBC database based on the shape of their CMDs and obtained spectra of their brightest members. We then derive parameters for these three clusters and proceed to a full characterisation in Sect.~\ref{section.results}. Finally, we discuss the limitations of photometric data.


\section{Identification of young clusters in the \textit{Gaia} CMD }
\label{sec:clusters}
   

 \begin{figure}
 \centering
  \resizebox{\columnwidth}{!}{\includegraphics[clip]{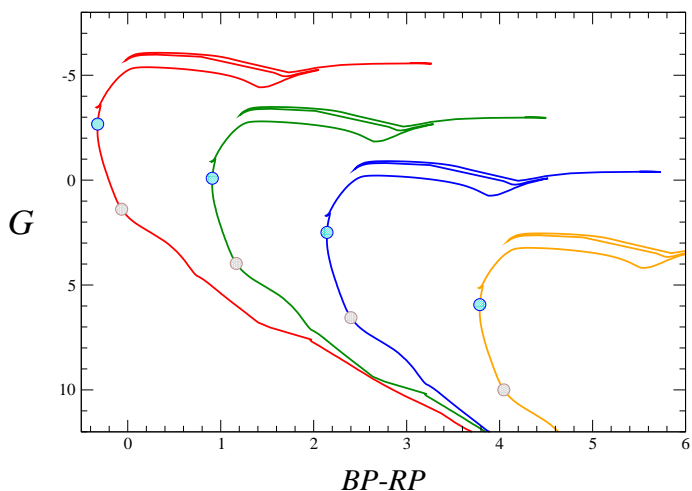}}
 \caption{Padova {\sc parsec} isochrones for 30~Ma and solar metallicity, affected by $A_{V}$ (from left to right) of 0, 3, 6, and 10~mag. The circles mark the position of stars with initial masses of $8\:\mathrm{M}_{\sun}$ (top, blue) and  $2.3\:\mathrm{M}_{\sun}$ (bottom, grey).}
  \label{isochrone}%
 \end{figure}
 

 \begin{figure*}
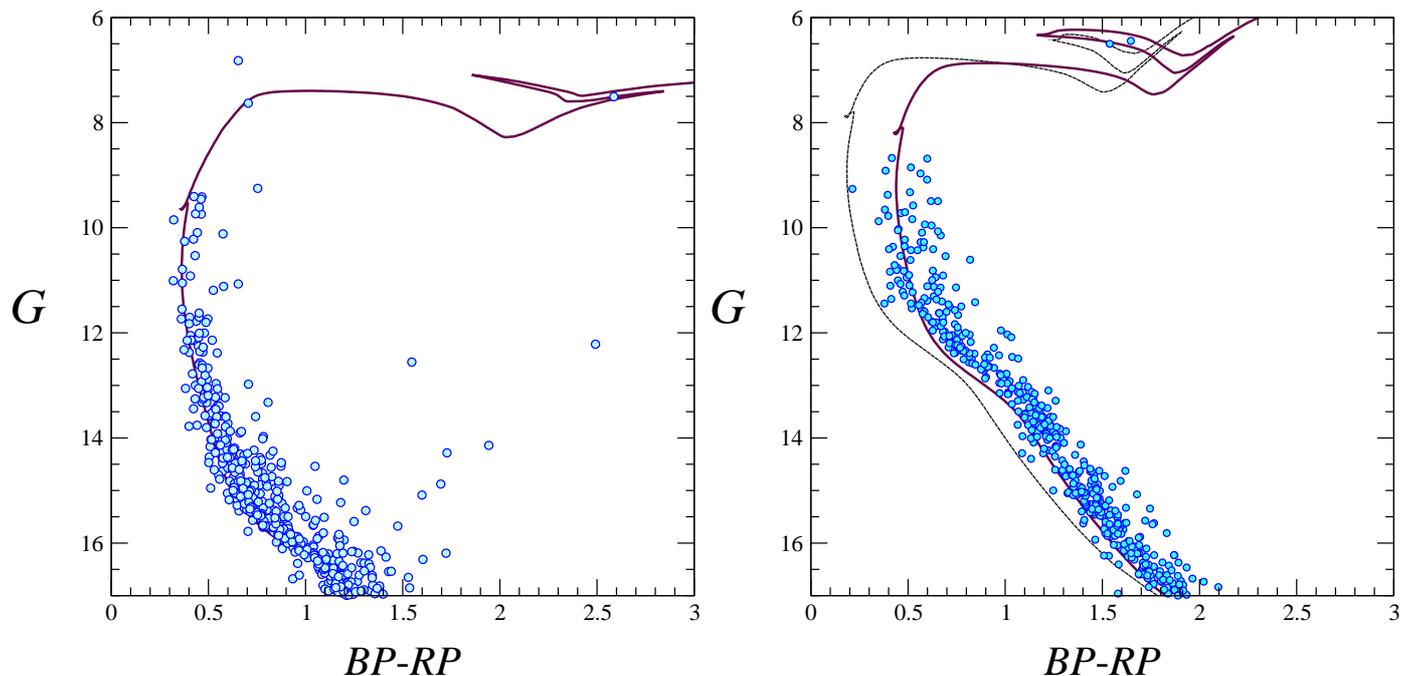

   \centering
   \resizebox{\textwidth}{!}{
 \includegraphics[width=\columnwidth, clip]{ngc457cmd.eps}
 \includegraphics[width=\columnwidth, clip]{ic4725cmd.eps}}
 \caption{Padova {\sc parsec} isochrone fits for the \textit{Gaia} DR2 CMDs of two well-studied young clusters. \textbf{Left :} NGC~457 fit with a 20~Ma solar metallicity isochrone. \textbf{Right :} IC~4725 fit with an 85~Ma solar metallicity isochrone (the dotted line shows the isochrone proposed by \citealt{cantat20}). See main text for details.}
              \label{compare}%
 \end{figure*}


For the sake of simplicity, in this work we will use mostly {\sc parsec} isochrones \citep{chen15} retrieved from the CMD~3.6 interface\footnote{\href{Padova}{http://stev.oapd.inaf.it/cgi-bin/cmd\_3.6}}. These isochrones correspond to non-rotating models with assumed solar metallicity $Z_{\sun}=0.015$. They were generated with the passbands from \cite{mapw18} and reddened automatically by assuming a \citet{cardelli} extinction law, which may be inadequate at high reddening values because of its polynomial nature, but is in most cases sufficiently accurate for moderate reddening.

To illustrate the shape of a young open cluster in \textit{Gaia} photometry, Fig.~\ref{isochrone} shows a 30~Ma isochrone affected by different amounts of extinction. As can be seen, the shape of the isochrone is not significantly affected by reddening. The isochrone presents three well-differentiated parts. The lower main sequence (MS) runs diagonally through the diagram, close to a straight line. An important feature of the \textit{Gaia} photometric system is that reddening moves stars roughly along the same line, so that the different isochrones end up running together. Because of this, determining the extinction -- and hence, age -- for an intermediate-age cluster with no evolved stars is almost impossible without additional information, as illustrated above by the case of UBC~94. The upper main sequence is much more vertical. In all the isochrones shown in Fig.~\ref{isochrone}, we have marked by circles the position of stars with $8\:\mathrm{M}_{\sun}$ (a B2\,V star at $Z_{\sun}$) and $2.3\:\mathrm{M}_{\sun}$ (around A0\,V). The almost vertical stripe corresponds thus to the position of B-type stars. The change in slope between these two sections, at the position of early A-type stars, is perhaps the most-easily identified feature in the MS of well-populated clusters. The post-MS evolution, with the almost horizontal track towards the red and the blue loop, is very well defined, but this region is, in most cases, very poorly populated. 

These features are illustrated in Fig.~\ref{compare}, where two well-studied and well-populated clusters are shown as an example of the shape of a young cluster CMD in \textit{Gaia} photometry. The left panel shows the CMD for the young cluster NGC~457. We have selected objects with membership probability $P\geq0.5$ in \citet{cantat18}. Given its moderate, uniform reddening and absence of significant foreground or background population \citep[cf.][]{negueruela17}, NGC~457 is a very good template for the shape of a real CMD. The exquisite quality of \textit{Gaia} photometry is manifest in the narrowness of the upper main sequence. The turn-offs of young open clusters are in general ill-defined, because of the combined effects of binary evolution (which generates blue stragglers) and a range of initial rotational velocities (which leads to broadened sequences). In fact, all the bright objects falling to the right of the sequence are known binaries or Be stars. We must point out that there are a few objects, all fainter than $G=12$, whose position is incompatible with the isochrone, which we consider astrometric interlopers (same astrometric parameters as the cluster, but not related)\footnote{The number of such interlopers will partly depend on the richness of the field, but also on how close the cluster astrometric parameters are to those of the surrounding field. See the case of UBC~103, discussed in Sect.~\ref{sec:interlopers}.}. These objects should be ignored. 

By assuming a distance modulus $DM = 12.2$, after \citet{cantat20}, the sequence is very well fit by a 20~Ma isochrone reddened by $A_{V}=1.5$, parameters that are directly comparable to those given by \citet{cantat20}, who finds the same age and a very slightly lower extinction of $A_{V}=1.42$. The age is determined by resorting to the location of the turn-off and the red supergiant HD~236697. We must note that two other cluster supergiants do not fit the isochrone. One is the B5\,Iab supergiant HD~7\,902, seen well above the isochrone in Fig.~\ref{compare}. The other is the F0\,Ia standard $\phi$~Cas, too bright to have \textit{Gaia} photometry. This discrepancy is discussed in \citet{negueruela17}, where a fit of Geneva \citep{ekstrom12} isochrones with $A_{V}=1.55$ and $DM =12.0$ (i.e., parameters compatible within errors) to $UBV$ photometry returns the same age estimate, demonstrating a rather high degree of consistency between different models and observed datasets. 

Not all real cluster sequences are as well defined as that of NGC~457. In many cases, differential reddening contributes as importantly as intrinsic effects to the broadening of the stellar sequence. As an example, the right panel of Fig.~\ref{compare} shows the older cluster IC~4725, which spans well over one degree on the sky in an area of variable obscuration. Accepting the distance modulus $DM = 9.2$ from \citet{cantat20}, the best fit is obtained by an 85~Ma isochrone reddened by $A_{V}=1.4$. These parameters are very different from the 110~Ma and $A_{V}=0.84$ given by \citet{cantat20}. Our younger age is determined by the location of the two yellow supergiants on the blue loop. The reason for the much higher extinction is obvious from Fig.~\ref{compare}, where we have also plotted the isochrone favoured by \citet{cantat20}. While our value of the reddening is driven by the bulk of the stars (and could indeed be even a little higher), the value given by the automated analysis represents a lower envelope for the set, determined by a few outliers with low extinction. In any event, the broadening of the upper main sequence still preserves the shape of the sequence and the inflection in its slope is perfectly noticeable, driving the right combination of distance modulus and reddening.

\section{Observations and astrometric data}
\label{section.obs_astrometry}

\subsection{Sample selection}


 \begin{figure*}[ht]
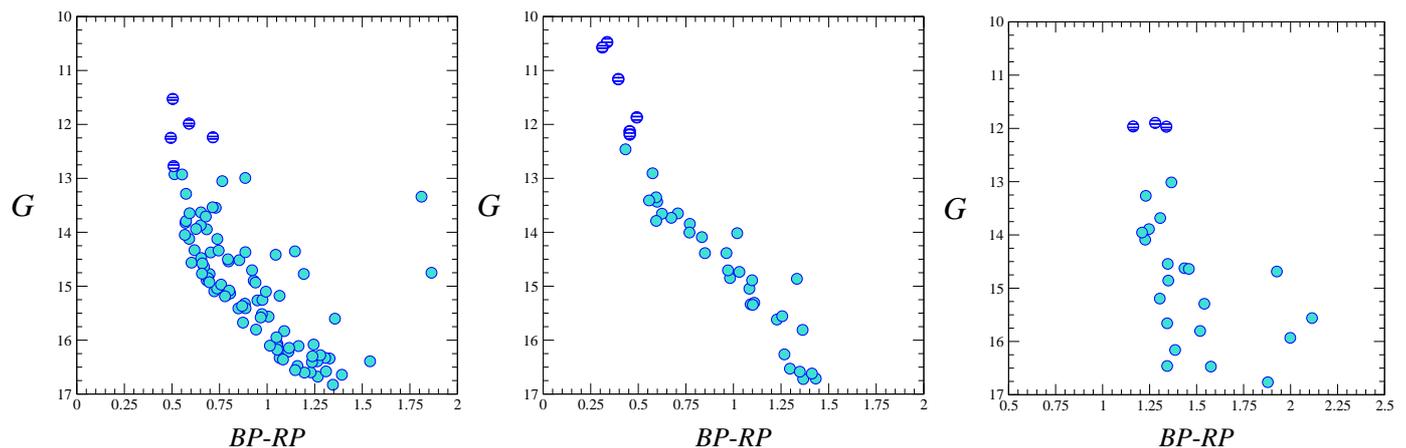

   \centering
   \resizebox{\textwidth}{!}{
 \includegraphics[width=0.32\columnwidth, clip]{ubc103_raw.eps}
 \includegraphics[width=0.32\columnwidth, clip]{ubc114_raw.eps}
 \includegraphics[width=0.32\columnwidth, clip]{ubc587_raw.eps}}
 \caption{Raw CMD diagrams for the three clusters observed, based on \textit{Gaia} photometry. From left to right, UBC~103, UBC~114 and UBC~587 (note the shift in the horizontal axis due to higher extinction). The large striped circles indicate the stars for which spectra were taken. Cluster members are from \citet{cantat20}. See main text for details.}
              \label{rawdata}%
 \end{figure*}


Three representative clusters were selected from the UBC database by resorting to visual inspection. The selection required the clusters to contain at least a moderate number of stars brighter than $BP= 13$ (so that spectra could be obtained with not very long exposure time), display an approximately vertical sequence in the CMD, which we have shown to be the identifying feature of young open clusters, and have good visibility in the early summer. 

The raw CMDs of the clusters selected are displayed in Fig.~\ref{rawdata}. The three clusters chosen cover a variety of situations  that can be found among the new cluster candidates. UBC~103 (left) is a well populated cluster, with a moderately broad main sequence, hinting at differential extinction. It is thus classified as an A category (easily identifiable) cluster by \citet{castro-ginard20}. Its slope inflection is clearly seen around $G\approx15$. UBC~114 (centre) is not so well populated, but has a narrow, very well-defined sequence, which still permits its classification as an A category cluster in \citet{castro-ginard20}. The slope inflection can be guessed slightly above $G\approx14$, but the number of stars above it is so small that the identification as a young cluster is uncertain. UBC~587 (right) is a very poorly populated cluster. The stellar sequence is broad and ill-defined, leading to its classification as a C category (doubtful) cluster candidate in \citet{castro-ginard20}. The sequence, however, is decidedly vertical over five magnitudes, suggesting a very young cluster.


 \subsection{Observations}
 \label{subsection.observations}
 
 Spectroscopic observations were taken using the Alhambra Faint Object Spectrograph and Camera (ALFOSC), attached to the 2.56~m Nordic Optical Telescope (NOT) at the Observatorio del Roque de los Muchachos (La Palma, Spain) of the Instituto de Astrof\'{\i}sica de Canarias. The instrument was equipped with CCD~\#14, a $2048\times2064$ e2v camera. The configuration used consists of grism \#18 and a $0\farcs75$ long slit. Grism \#18 provides a wavelength range of 345\,--\,535\,nm, with a central wavelength at 436\,nm and an approximate resolving power of $R\sim2\,000$.


\begin{table*}[ht]
\caption{Some basic parameters of the stars observed in the field of UBC~103, UBC~114 and UBC~587, together with the exposure times used.}
\label{tab:observations}
\centering
\begin{tabular}{l c c c c c c c}
\hline\hline
\noalign{\smallskip}
ID& Name & R.A. & Dec. & $BP$ & $RP$ & Exposure & Spectral\\
&&&&(mag) &(mag) & Time & Type\\
\noalign{\smallskip}
\hline\noalign{\smallskip}

\multicolumn{8}{c}{UBC103} \\\noalign{\smallskip}
1 & TYC 5125-2315-1 & 18:42:29.5 & $-$06:34:16 & 11.70 & 11.20 & 360 & B4\,III \\
2 & $-$             & 18:42:33.8 & $-$06:37:39 & 12.20 & 11.60 & 440 & B5\,IV \\
3 & $-$             & 18:42:43.6 & $-$06:39:56 & 12.50 & 11.79 & 400 & B5\,V \\
4 & TYC 5125-2466-1 & 18:42:39.1 & $-$06:35:20 & 12.42 & 11.92 & 440 & B6\,IV \\
5 & $-$             & 18:42:31.7 & $-$06:35:15 & 12.92 & 12.42 & 590 & B5\,V \\

\noalign{\smallskip}
\hline
\noalign{\smallskip}

\multicolumn{8}{c}{UBC114} \\\noalign{\smallskip}
1 & HD 177631      & 19:05:32.1 & $-$00:16:15 & 10.59 & 10.26 & 200 & B2\,V   \\
2 & TYC 5128-116-1 & 19:05:37.8 & $-$00:24:29 & 10.68 & 10.37 & 140 & B2.5\,V \\
3 & TYC 5128-702-1 & 19:06:11.9 & $-$00:30:05 & 11.29 & 10.90 & 300 & B5\,V  \\
4 & $-$            & 19:05:50.7 & $-$00:22:35 & 12.28 & 11.83 & 450 & B8.5V   \\
5 & $-$            & 19:05:46.5 & $-$00:25:32 & 12.34 & 11.89 & 470 & B8\,V   \\
6 & TYC 5128-593-1 & 19:05:16.3 & $-$00:35:22 & 12.04 & 11.55 & 440 & B9\,V   \\ 

\noalign{\smallskip}
\hline
\noalign{\smallskip}

\multicolumn{8}{c}{UBC587} \\\noalign{\smallskip}
1 & BD $+43\degr$3710 & 20:45:34.7 & +43:32:27 & 10.24 &  8.21 & 180 & B1\,Ia \\
2 & BD $+43\degr$3714 & 20:46:46.3 & +43:47:34 & 10.79 &  9.39 & 240 & B0\,II \\
3 & ALS 11533         & 20:46:13.8 & +43:44:59 & 11.53 & 10.22 & 270 & O9.7\,V \\
4 & $-$               & 20:46:18.3 & +43:35:04 & 12.45 & 11.29 & 440 & B1\,V \\
5 & TYC 3178-867-1    & 20:46:18.6 & +43:44:13 & 12.45 & 11.17 & 410 & B0.5\,IV \\
6 & $-$               & 20:46:48.3 & +43:47:30 & 12.56 & 11.22 & 410 & B0.7\,V \\ 

\noalign{\smallskip}
\hline

\end{tabular}
\end{table*}


 With this configuration, we gathered 19 spectra of different sources selected from the three clusters described above. Observations were taken as part of the NOT Fast-Track program during two consecutive bright nights on June 6 and 7, 2020. A summary of the target properties and exposure times used can be found in Table~\ref{tab:observations}. Weather conditions were very good, with a seeing of $\sim0\farcs9$ measured on the acquisition images. We aimed at a signal-to-noise ratio of $\sim$150 in the 500\,--\,600~nm range.

 The reduction was carried out using Python programming language and standard tasks within the IRAF software environment. The full reduction of the spectra included bias and flat-field correction, cosmic-ray removal, sky subtraction, wavelength calibration by using Th-Ar arc-lamps, and flux calibration using spectra of standard stars observed with the same setup during the same nights. 
 
 Spectral classification was carried out by following the traditional classification procedure through a comparison to a new, large grid of standard stars \citep[and in preparation]{negueruela18}, degraded to the same resolution as the target spectra.


 \subsection{Astrometric data}
 \label{subsection.astrometry}

 As a first step, astrometric data from \citet{castro-ginard20} were downloaded from the CDS/\textit{Vizier} server. These include information on parallaxes ($\varpi$) and proper motions ($\mu_{\alpha}$, $\mu_{\delta}$) from {\em Gaia} DR2 \citep{2018A&A...616A...1G, 2018A&A...616A...2L} for the individual members of each cluster as determined by their algorithm. 
 We then used the Topcat\footnote{\href{Topcat}{http://www.star.bris.ac.uk/~mbt/topcat/}} Virtual Observatory tool \citep{TOPCAT2005} to crossmatch the coordinates of all these stars with the {\em Gaia} EDR3 database \citep{2021A&A...649A...1G}. For the crossmatches, we used a radius threshold of 1\,arcsecond. The EDR3 data for this set of likely members were used to recalculate average cluster values in EDR3.
 After this, we used Topcat again to retrieve all stars in {\em Gaia} EDR3 matching two conditions: firstly, they must be within half a square degree of the average coordinates of the stars in each cluster; secondly, they must be within 2\,$\sigma$ from the average value of $\varpi$, $\mu_{\alpha}$ and $\mu_{\delta}$ in EDR3 for their corresponding cluster. The associated errors of each individual value were not considered. This procedure may leave out some stars that are in fact members, but have divergent astrometric values with large uncertainties\footnote{In fact, the uncertainties provided in the database are certainly underestimations for the parallax values. See \citet{maiz21} for the best way to estimate realistic parallax uncertainties.}. However, this does not affect the average cluster values that we are deriving \citep[see, for example,][for a procedure to detect all likely members and slow ejections from a given cluster]{maiz_villa1}.


\section{Results}
\label{section.results}
  

In order to test the validity of the cluster parameters and cluster membership obtained with the (semi)automatic procedure of \citet{castro-ginard20}, we used both the spectroscopic and astrometric information available (see Sect.~\ref{subsection.observations}). 
We found that some stars that were considered cluster members by \citet{castro-ginard20}, based on their \textit{Gaia} DR2 parameters, lie outside 3\,$\sigma$ of the new average \textit{Gaia} EDR3 values in $\varpi$, $\mu_{\alpha}$ and $\mu_{\delta}$. We removed them and used the remaining stars to recalculate these averages. Once the new EDR3 cluster parameters were thus fixed, we selected all \textit{Gaia} EDR3 sources within 2\,$\sigma$ from these average values in all astrometric parameters. In the case of $\varpi$, we first corrected the values from zero-point offset by using the procedure described in \citet{2021A&A...649A...4L}\footnote{We used the algorithm provided by these authors at
{\tt https://gitlab.com/icc-ub/public/gaiadr3\_zeropoint}}.

From the sources selected, we took those brighter than $G=17.4$ with an on-sky distance $<0\fdg25$ deg from the nominal cluster location in \citet{castro-ginard20}. This was felt to be a sensible compromise between the need to probe a larger spatial extent than determined by \citet{castro-ginard20} and the danger of including other nearby stellar systems. Finally, we filtered out the sources with Renormalised Unit Weight Error (RUWE) $>1.4$ or with an error in parallax higher than 33\%. This gave us a final list of stars whose parameters are compatible with cluster membership out to a large distance from their putative centres.

In a final step, we built the \textit{Gaia} CMD -- $G$ vs. $(BP-RP)$ -- with all the stars selected through this procedure. We marked the members selected by \citet{castro-ginard20}, unless they had been removed, and compared their position in the diagram with the rest of the stars that have compatible EDR3 astrometric values, but had not been selected as members by their algorithm.
In each of the clusters, we estimated the age of the population and the extinction ($A_{V}$) by visually fitting the isochrones, paying special attention to the position of the stars that we had observed spectroscopically, and exploring the combination of both parameters. For the sake of completeness, in addition to the Padova (CMD 3.6) isochrones, we also downloaded MIST isochrones from their website\footnote{\href{MIST}{http://waps.cfa.harvard.edu/MIST/interp\_isos.html/}} (version 1.2). In all cases, the distance modulus assumed for the fit corresponds to the distance obtained by simple inversion of the average \textit{Gaia} EDR3 parallax, calculated as described at the start of this section. We then explored if the observed spectral types of the brightest members were compatible with their location in the CMD. 
A summary of the results derived from the isochrone fitting for each of the clusters compared to those from \citet{cantat20} is included in Table~\ref{tab:summary}.

\begin{table}[ht]
\caption{Summary of parameters of UBC~103, UBC~114 and UBC~587 derived from the isochrone fitting compared with the results from \citet{cantat20}.}
\label{tab:summary}
\centering
\begin{tabular}{c c c c}
\hline\hline
\noalign{\smallskip}
& Age & A$_{V}$ & DM \\
&(Ma) &(mag) & (mag) \\
\noalign{\smallskip}
\hline\noalign{\smallskip}

\multicolumn{4}{c}{UBC103} \\\noalign{\smallskip}
Cantat-Gaudin & 22 & 1.4 & 12.37 \\
This work     & $70\pm10$ & 1.6 & 12.41 \\

\noalign{\smallskip}
\hline
\noalign{\smallskip}

\multicolumn{4}{c}{UBC114} \\\noalign{\smallskip}
Cantat-Gaudin & 50 & 1.0 & 10.9 \\
This work     & $<30$ & 1.3 & 10.84 \\

\noalign{\smallskip}
\hline
\noalign{\smallskip}

\multicolumn{4}{c}{UBC587} \\\noalign{\smallskip}
Cantat-Gaudin & 14 & 3.1 & 12.2 \\
This work     & $>3$ \& $<10$ &$>3.3^{1}$ & 12.44 \\

\noalign{\smallskip}
\hline
\noalign{\smallskip}

\end{tabular}
\tablefoot{$^{1}$ Although many cluster members form a sequence that can be fitted by $A_V=3.3$, many others have obviously higher exinction.}
\end{table}
 

 \subsection{UBC~103}
 \label{subsection.res_ubc103}
 
\begin{figure*}[!t]
 \centering
 \includegraphics[width=1\textwidth]{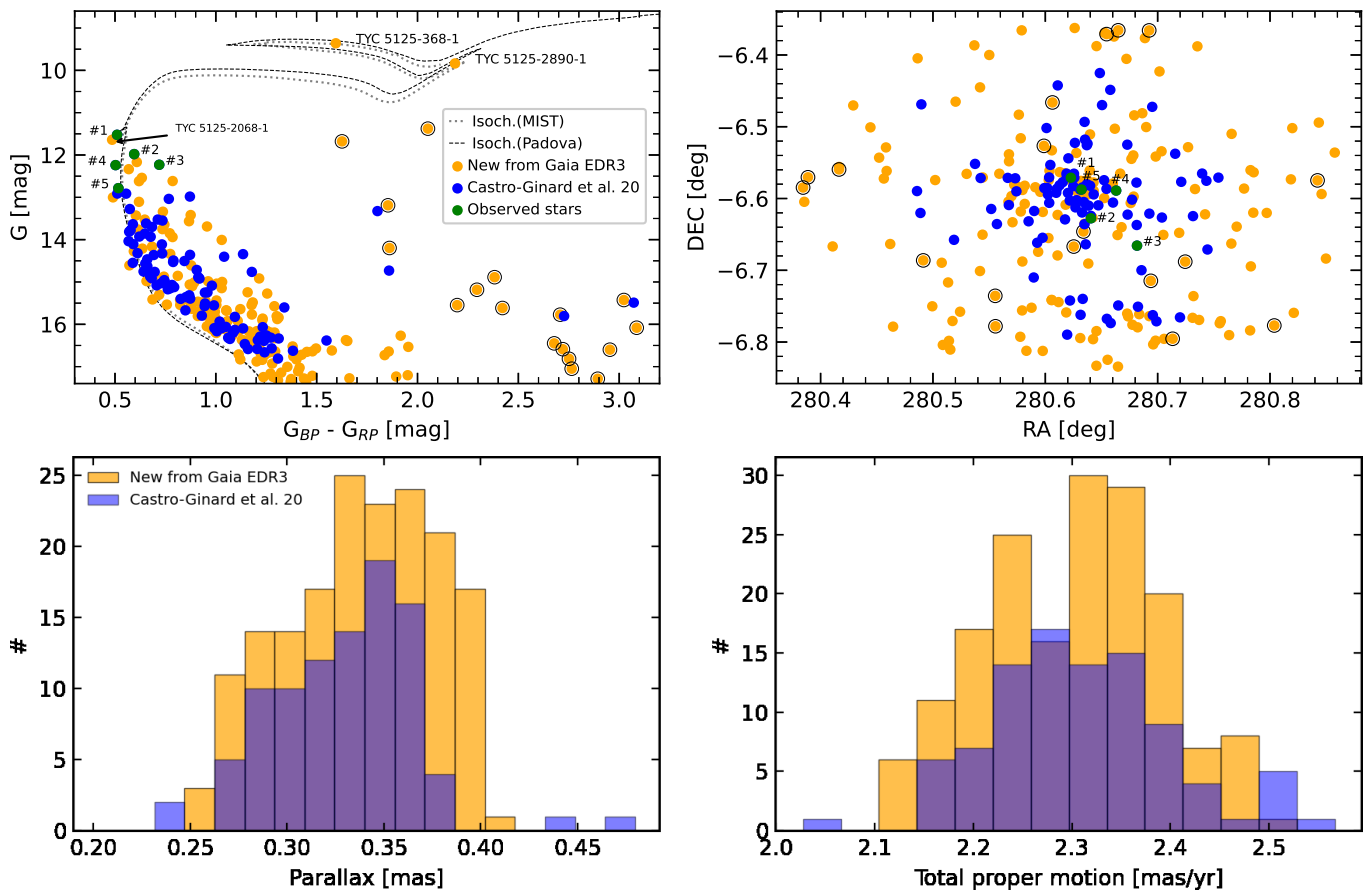}
 \caption{Summary figure for UBC~103. \textit{Top two figures:} Colour-magnitude diagram (left) and sky coordinates of the sources (right). \textit{Bottom two figures:} Histograms of the parallax (left) and total proper motion (right). In \textit{all figures}, blue circles or bins indicate cluster stars listed in \citet{castro-ginard20}, here represented by their {\em Gaia} EDR3 data; filled green circles are the observed stars; filled orange circles or bins indicate other stars found in {\em Gaia} EDR3 within 2\,$\sigma$ from the average cluster values (see Sect.~\ref{section.results}). Additionally, empty black circles indicate a potential separate population of stars (see Sect.~\ref{subsection.res_ubc103}); the gray dashed and dotted lines correspond to the best fitted isochrones from Padova and MIST databases, respectively, both for a value of 70~Ma and $A_{V}=1.6$.}
 \label{figure.ubc103}
 \end{figure*}

 UBC~103 is a well populated cluster. It lies on the edge of the Scutum Star Cloud (at $l=26.0$, $b=-1.0$) and is thus projected on an area of very high stellar density. Its CMD (see Fig.~\ref{figure.ubc103}, top-left panel) is suggestive of a moderate degree of differential reddening. The mean parameters found by \citet{castro-ginard20} are $\varpi=0.28\pm0.03\:$mas, $\mathrm{pm} =(-0.40\pm0.09$,$-2.27\pm0.09)\:$mas\,a$^{-1}$. The automated procedure of \citet{cantat20} finds a $DM=12.37$, $A_V=1.4$ and an age of 22~Ma. It is not possible, however, to obtain a good fit with these parameters, as the extinction needs to be higher for any sensible age. 
 In {\em Gaia} EDR3, the average cluster values, after applying a zero-point offset correction to the parallax, are $\varpi=0.33\pm0.06\:$mas, $\mathrm{pm} =(-0.38\pm0.08$,$-2.28\pm0.10)\:$mas\,a$^{-1}$.

\begin{figure}
 \centering
  \resizebox{\columnwidth}{!}{\includegraphics[clip, angle=-90]{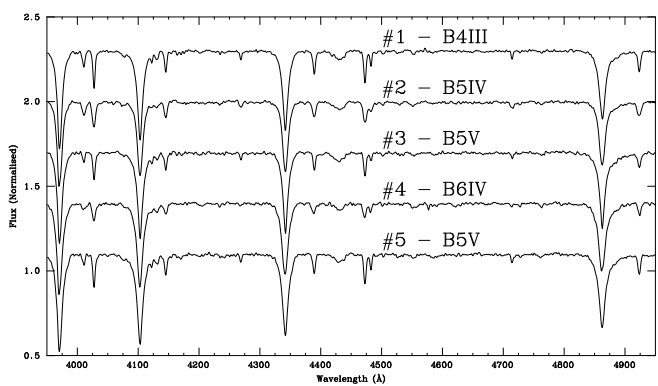}}
 \caption{Classification spectra of the brightest members of UBC~103 from \citet{castro-ginard20} ordered by spectral type.}
  \label{spectra103}%
 \end{figure}

 We observed the five brightest members according to \citet{castro-ginard20}. Their spectra are displayed in Fig.~\ref{spectra103}. The spectral types have been derived by assuming that they are all single stars. However, star \#4 is almost certainly a binary, based on its broad asymmetric lines. Aside from that, all their spectral types (Table~\ref{tab:observations}) are very similar, with an apparent turn-off at B5\,V and some slightly evolved stars. These spectral types are inconsistent with the young age proposed by \citet{cantat20}, as a turn-off at B5\,V occurs at ages in the 60 to 100~Ma range at solar metallicity. 
 
Amongst the EDR3 new candidate cluster members, we find three catalogued bright objects. One of them, TYC~5125-2068-1 was photometrically classified as a B5 giant by \citet{roslund63}. It has very similar magnitude and colour to the brightest member for which we have a spectrum, TYC 5125-2315-1 (B4\,III). Two other bright stars with redder colours occupy a position in the CMD compatible with cluster membership (top-left panel of Fig.~\ref{figure.ubc103}). They are TYC~5125-368-1, whose location in the CMD suggests it is a yellow supergiant, and TYC~5125-2890-1, which occupies a position expected for a red supergiant. There are three strong reasons to believe that these two objects are cluster evolved members:
 \begin{enumerate}
    \item Their position in the 2MASS CMD ($(J-K_{\mathrm{S}}$) vs. $K_{\mathrm{S}}$) is also compatible with the cluster isochrone (see Fig.~\ref{fig:IRCMD}).
    \item Their EDR3 radial velocities are $44.6\pm0.2$ and $45.6\pm0.1\:\mathrm{km}\,\mathrm{s}^{-1}$, confirming that they belong to the same population. The only other red star with similar brightness has an incompatible value, and velocities around $45\:\mathrm{km}\,\mathrm{s}^{-1}$ are not particularly common in this field\footnote{\citet{castro-ginard20} quote a RV of $-3.99\:\mathrm{km}\,\mathrm{s}^{-1}$ for UBC~103. This is based on the RV of a single star that, given its position in the CMD, is not a cluster member. In fact, it is part of the redder population with similar proper motions discussed below}.
    \item Stars of such brightness at this distance are in all likelihood supergiants, and such objects are much more likely to be associated to the young cluster.
 \end{enumerate}


 These two candidate supergiants allow a firm determination of the cluster age. In Fig.~\ref{figure.ubc103}, we show the fit of a 70~Ma isochrone reddened with $A_{V}=1.6$, which gives a rather good fit for the distance modulus $DM=12.4$ derived from the \textit{Gaia} parallax\footnote{Nevertheless, we have to note that, in the absence of a \textit{Gaia} parallax, photometry alone would have favoured a distance modulus $DM=12.0$, implying a slightly older age. It is difficult to estimate how significative this difference is, as, at this large distance, the small dispersion of the parallax values ($0.33\pm0.06$) would translate into a large range of distance modules ($12.4\pm0.4$) and a huge range of distances ($3.0^{+0.7}_{-0.4}$~kpc), if directly propagated.}. Without the evolved stars, it is possible to obtain good fits for younger ages with slightly higher extinctions. The observed spectral types at the top of the main sequence, however, are incompatible with ages much younger than $\sim 60\:$Ma. Ages up to 100~Ma would also be compatible with a turn-off at B5\,V (cf. the case of IC~4725). 
 
 \subsubsection{A population of interlopers}
 \label{sec:interlopers}
 
 The search for new members among {\em Gaia} EDR3 sources out to 15 minutes from the centre adds a substantial population to the main sequence (top-left panel of Fig.~\ref{figure.ubc103}), but also includes a second sequence at much redder colours -- about one magnitude in $(BP-RP)$ -- that was already hinted at in the sample of \citet{castro-ginard20}. The majority of these redder sources lie at large distances from the main cluster concentration (top-right panel of Fig.~\ref{figure.ubc103}), suggesting that they represent a different population. To confirm this, we cross-matched the selection with the 2MASS database. The field is very crowded, and the possibility of spurious matches is high. We therefore kept only 2MASS sources with distances smaller than $0\farcs2$ from the corresponding DR3 sources. Also because of crowding, only a small fraction of sources have good flags and acceptable photometric errors. Although we were lax with the photometric precision required, allowing errors up to $0.08$~mag in each of the three ($JHK_{\mathrm{S}}$) bands, we are left with only 75 sources, many of them with moderately large errors.

 \begin{figure}
 \centering
 \resizebox{\columnwidth}{!}{\includegraphics[clip]{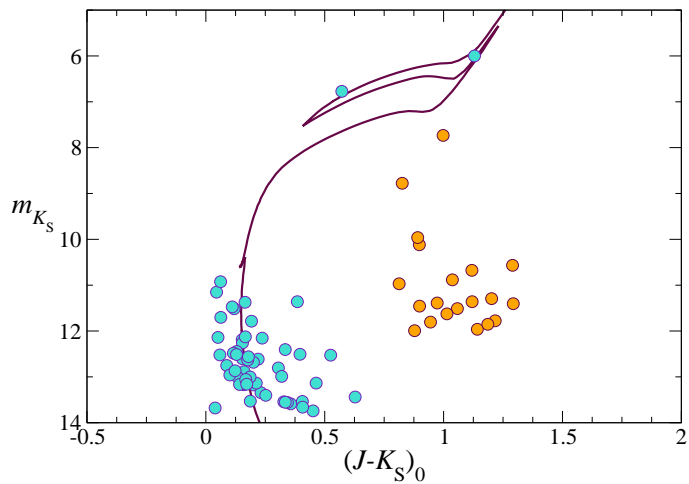}}
 \caption{Infrared CMD for the stars selected as astrometric members of UBC~103. The blue circles are the 2MASS counterparts of objects that fall in the cluster sequence and the two objects that we interpret as cluster supergiants. The orange circles are the counterparts to the objects that present redder colours in the \textit{Gaia} CMD. They are also clearly displaced to redder colours, as well as to brighter magnitudes, in this plot, indicating a separate population. 
              \label{fig:IRCMD}}%
    \end{figure}
    
The CMD for this sample of objects is shown in Fig.~\ref{fig:IRCMD}, where we have separated the stars that fall on the cluster sequence in the \textit{Gaia} CMD from those in the second sequence at redder colours. We see that these objects also have much redder 2MASS colours. In addition, the objects in the red sequence are much brighter in $K_{\mathrm{S}}$ than the stars on the blue sequence of the same $G$, confirming that they are intrinsically red stars. For the brightest objects, which have accurate 2MASS photometry, we calculated their $Q_{\mathrm{IR}}$ reddening-free indices. While the brightest objects in the blue sequence have $Q_{\mathrm{IR}}\la0.1$, typical of early-type stars, the objects in the red sequence have $Q_{\mathrm{IR}}\ga0.2$, again indicating that they are intrinsically redder objects \citep{ns07}. The only possible interpretation is the presence of a different population with the same astrometric parameters as cluster members. We may speculate with a population of red giants, which would be compatible with the red colours and bright infrared magnitudes.

\subsubsection{A twin cluster and an association}

 Even if we remove this second population, there are many stars whose astrometric and photometric parameters are compatible with membership at high distances from the nominal cluster centre (top-right panel of Fig.~\ref{figure.ubc103}). If we increase the radius of our search, rather than finding a decrease in the density of candidate members in all directions, we come across a very strong concentration located $\sim22\arcmin$ to the north of UBC~103, coincident with the position of the known open cluster NGC~6883. Unfortunately, this cluster is not included in the catalogue of \citet{cantat20}. In Appendix~\ref{app:ngc}, we calculate its astrometric parameters, which are fully compatible with those of UBC~103, making them a couple of twin clusters.
 
 The distribution of astrometric members in the sky suggest that NGC~6883 and UBC~103 are the main concentrations of a rather large population spread over a large area, which extends well beyond the $15\arcmin$ covered by our initial search. To estimate its size, we used the Virtual Observatory tool Clusterix 2.0 \citep{balaguer20}, an interactive web-based application that calculates the grouping probability of a list of objects by using proper motions and the non-parametric method proposed by \citet{cabreracano90} and described in \citet{galadi98}. We run Clusterix on circles of increasing radius around UBC~103, restricting the search to stars brighter than $G=16$ for ease of computation. We find significant numbers of stars with compatible proper motion and parallax out to $40\arcmin$, especially towards the north and northeast of NGC~6883 and to the south of UBC~103. There are few stars belonging to this population to the west, but this is in all likelihood due to the presence of dark clouds, which render the density of bright stars much lower.  Towards the east, we move deep into the Scutum Star Cloud, where crowding becomes an issue, even for \textit{Gaia}. We cannot continue the search at larger radii, as we run into the very massive open cluster Valparaiso~1 \citep[= UBC~106; $d\approx 2.3\:$kpc, $\tau\approx75\:$Ma; ][]{negueruela21}, whose centre is located slightly more than $1\degr$ due North from UBC~103, and its larger population starts to dominate the frequency distribution in the vector point diagram.  
 
 \begin{figure}
 \centering
 \resizebox{\columnwidth}{!}{\includegraphics[clip]{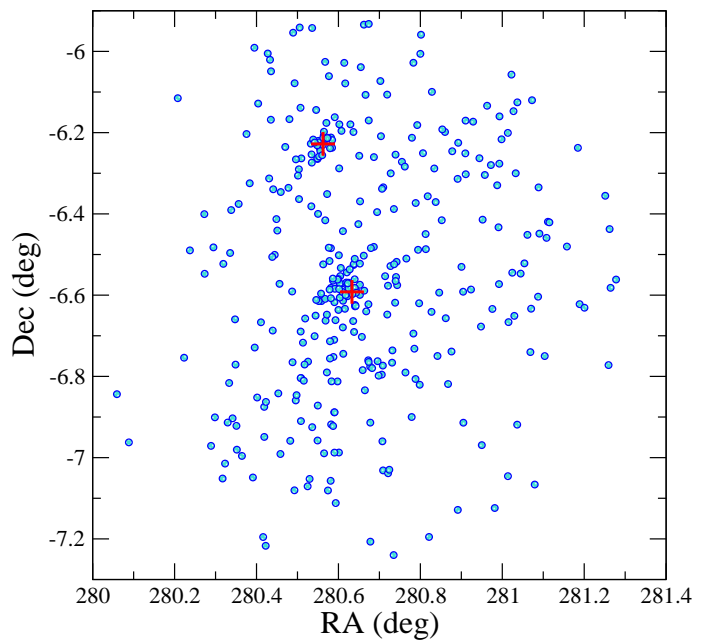}}
 \caption{Spatial distribution of likely members of the stellar association surrounding UBC~103 out to $40\arcmin$ from its nominal centre. High-probability members in the space of proper motions found by Clusterix~2.0 have been pruned by parallax and position in the CMD (only stars with $10\leq G\leq 16$ are included). The red crosses mark the nominal position of UBC~103 (in the centre of the search area) and the position of NGC~6883 found in Appendix~\ref{app:ngc}. Note the lack of members to the West, most likely due to intervening dark clouds causing heavy extinction.
              \label{fig:extent}}%
    \end{figure}
    
 The results of this search can be seen in Fig.~\ref{fig:extent}. We took EDR3 data for stars in the $10\leq G\leq16$ range within a circle of radius $40\arcmin$ around the nominal position of UBC~103, and then selected the objects identified by Clusterix 2.0 as high probability members of a homogenous population in proper motion space. Afterwards, we pruned the sample by requiring a parallax consistent with that of UBC~103 and a position in the CMD compatible with the sequences of UBC~103 or NGC~6883. The result is a list of 358 stars, very strongly concentrated towards UBC~103, but including several other clumps, the most populous of which is NGC~6883. By comparison, there are 34 objects within $5\arcmin$ of the centre of NGC~6883 and 58 objects within $5\arcmin$ of the nominal position of UBC~103.
 
 Extending, for completeness, the search to brighter stars, up to $G=9$, recovers the two likely supergiant members of UBC~103 mentioned above and adds a single, very interesting source: the 12.9~d Cepheid Z~Sct. This object, located close to the northernmost edge of our search area, has astrometric parameters fully consistent with membership and average photometric colours almost identical to those of TYC~5125-368-1 (it is about 0.1~mag brighter in all bands). Its \textit{Gaia} DR3 RV is $53.3\pm3.6\:\mathrm{km}\,\mathrm{s}^{-1}$, where the large dispersion is due to intrinsic variability. \citet{anderson16}, however, calculate a systemic velocity of $29.9\:\mathrm{km}\,\mathrm{s}^{-1}$, which is inconsistent with the RVs of the two members of UBC~103. The membership of this object is therefore not secure.
 
 At a distance $\sim3.0\:$kpc, a diameter of $80\arcmin$ corresponds to a $\sim70\:$pc span. The existence of such large stellar complexes in this direction is not unexpected. In fact, recently \citet{kuhn21} identified some larger complexes of star forming regions towards the inner Galaxy, including a huge high-pitch structure extending over almost 1~kpc between $l=4\degr$ and $18\degr$. In the direction to UBC~103 ($l\sim26\degr$), the recently identified open cluster UBC~1035, located about $1\degr$ to the west, has the same proper motions, parallax and estimated age as Valparaiso~1, which is $\sim80\arcmin$ away. The projection of several large clusters with ages between $\sim50$ and $\sim100$~Ma and slightly different distances could indicate that a structure similar to that found by \citet{kuhn21} produced abundant star formation in the past.

 \subsection{UBC~114}
 \label{subsection.res_ubc114}
 
  UBC~114 is a poorly populated cluster at a relatively low distance. The grouping appears projected on top of a dark cloud, at $l=32\fdg6$, $b=-2\fdg7$. The mean parameters found by \citet{castro-ginard20} are $\varpi=0.66\pm0.03\:$mas, $\mathrm{pm} =(-1.50\pm0.10$,$-1.88\pm0.09)\:$mas\,a$^{-1}$. The automated procedure of \citet{cantat20} finds a $DM=10.91$, $A_V=1.0$ and an age of 50~Ma. 
  In {\em Gaia} EDR3, the average values for the same stars, after applying a zero-point offset correction for the parallax, are $\varpi=0.68\pm0.03\:$mas, $\mathrm{pm} =(-1.48\pm0.06$,$-1.95\pm0.10)\:$mas\,a$^{-1}$.
  
\begin{figure}[b]
 \centering
  \resizebox{\columnwidth}{!}{\includegraphics[clip, angle=-90]{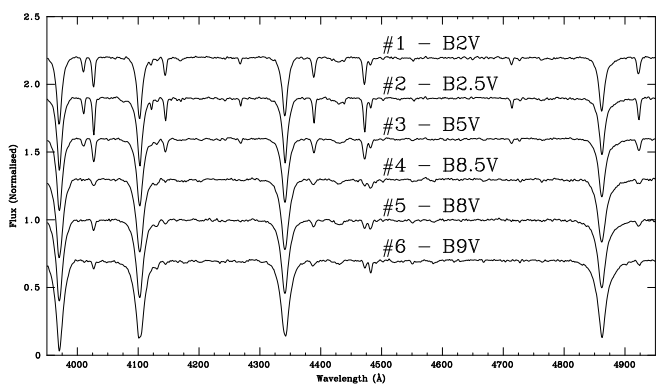}}
 \caption{Classification spectra of the brightest members of UBC~114 from \citet{castro-ginard20} ordered by spectral type.}
  \label{spectra114}%
 \end{figure}

 We observed the six brightest members from the list of \citet{castro-ginard20}, whose classification spectra are displayed in Fig.~\ref{spectra114}. Their spectral types (Table~\ref{tab:observations}; mid panel) cover most of the range of B-type dwarfs. Again, all the spectral types assume that the stars are single. Star \#5 is very likely a binary, given its broad, flat-bottomed \ion{He}{i} lines. Star \#6 is also possibly a binary, as it is brighter than stars of earlier spectral type. 

\begin{figure*}[h!]
 \centering
 \includegraphics[width=1\textwidth]{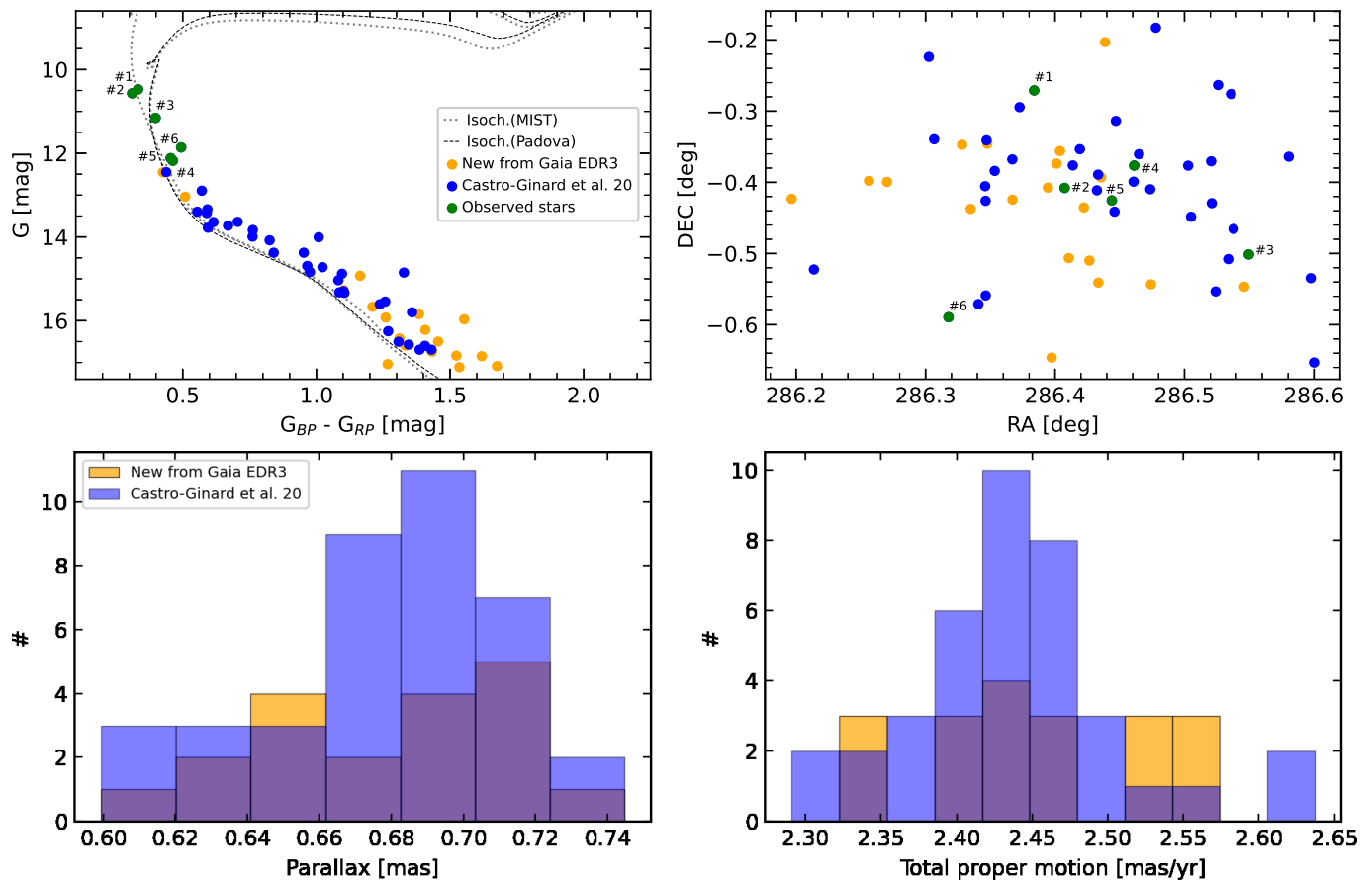}
 \caption{Summary figure for UBC~114. \textit{Top two figures:} Colour-magnitude diagram (left) and sky coordinates of the sources (right). \textit{Bottom two figures:} Histograms of the parallax (left) and total proper motion (right). See Fig.~\ref{figure.ubc103} for a description of the symbols and colours. The two grey lines next to each other are isochrones of 100~Ma with $A_{V}=1.2$ (the dashed line is a CMD 3.6 isochrone, while the dotted line is from the MIST database). The dotted line to the left of the others  corresponds to a 26~Ma isochrone with $A_{V}=1.3$, taken from MIST. Data seem to favour the younger age.}
 \label{figure.ubc114}
 \end{figure*}

The search in EDR3 only adds faint members to the cluster sequence, confirming that there is no obvious contaminating population. The upper main sequence is very poorly populated, and there are no evolved stars to fix the cluster age. The observed distribution of spectral types allows the cluster to be as young as wished, with an upper limit on the age given by the turn-off age of a B2\,V star, around 25~Ma. As an example, in Fig.~\ref{figure.ubc114} we plot an isochrone of 26~Ma and $A_{V}=1.3$, where the \textit{Gaia} parallax provides $DM=10.8$.
  
 Alternatively, one could consider that the turn-off regions of the CMDs of young clusters are in many cases dominated by the effects of binary interaction and assume that the most luminous stars are blue stragglers of some sort. In support of this interpretation, the EDR3 astrometric values for star \#3 are very different from those of DR2, and the star is no longer selected as a cluster member, given that its $\mu_{\delta}$ ($-2.34\pm0.02\:$mas\,a$^{-1}$) is significantly different from the cluster average. If we assume that this object is not a member, then we find a huge gap of almost two magnitudes between the two stars at the top of the sequence and the faintest stars for which we have spectra, at types B8\,--\,9. The presence of such a gap may lead us to suspect that the actual age of the cluster is given by these later type stars, because the sequence is only continuous from that point, implying that the two early-type stars are some sort of blue stragglers.
 
This scenario, however, is unlikely, because a B2\,V star has a mass of $\sim8\:\mathrm{M}_{\sun}$ and thus cannot be formed even by the direct merger of two B8\,V stars ($\sim3.3\:\mathrm{M}_{\sun}$; \citealt{harmanec88}). Therefore, we would have to assume that the blue stragglers have formed by the interaction of stars more massive than any that we see, which again requires a younger age. Moreover, turning back to star \#3, if we consider that its parallax and $\mu_{\alpha}$ are fully consistent with the cluster average, it is still possible to assume that the object is connected to the cluster. Given that it lies about 9~arcmin south of the nominal centre (see Fig.~\ref{figure.ubc114}, right-top panel), this object may be a slow runaway in the process of being ejected. Therefore, we consider that this is indeed a very young cluster. Finally, we note that, even for older ages, the $A_{V}=1.02$ value derived by \citet{cantat20} is \textit{again} too low to permit a good fit to the bulk of the stars. Our fits require $A_{V}=1.3$ for ages younger than $\sim30\:$Ma and $A_{V}=1.2$ for older ages, up to $\sim150$~Ma.


 \subsection{UBC~587}
 \label{subsection.res_ubc587}
 
 UBC~587 is an extreme case, as its raw CMD does not show evidence of any distinctive shape. The distribution of stars in the CMD, however, suggests a vertical stripe affected by strong differential reddening. The mean parameters found by \citet{castro-ginard20} are $\varpi=0.28\pm0.02\:$mas, $\mathrm{pm} =(-2.87\pm0.08$,$-4.84\pm0.13)\:$mas\,a$^{-1}$. The high distance implied further suggests a young open cluster. The automated procedure of \citet{cantat20} finds a $DM=12.2$, $A_V=3.1$ and an age of 14~Ma. 
 In {\em Gaia} EDR3, the average values for the same stars, after applying a zero-point offset correction for the parallax, are $\varpi=0.33\pm0.04$~mas, $\mathrm{pm}=(-2.85\pm0.09$,$-4.86\pm0.13)\:$mas\,a$^{-1}$. 
 
 \begin{figure}[]
 \centering
  \resizebox{\columnwidth}{!}{\includegraphics[clip, angle=-90]{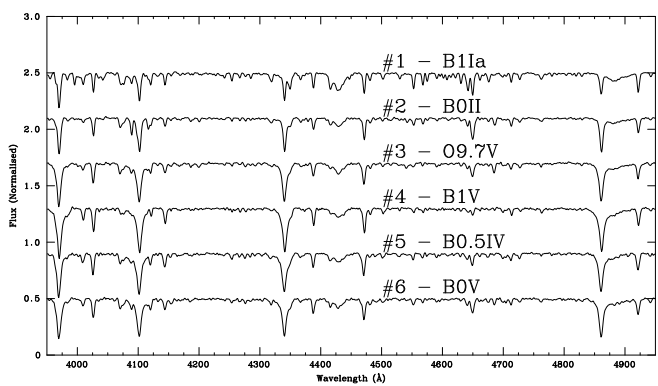}}
 \caption{Classification spectra of the three brightest members of UBC~587 according to \citet{castro-ginard20} and three brighter catalogued OB stars in the area.}
  \label{spectra587}%
 \end{figure}
 
 \begin{figure*}[!]
 \centering
 \includegraphics[width=1\textwidth]{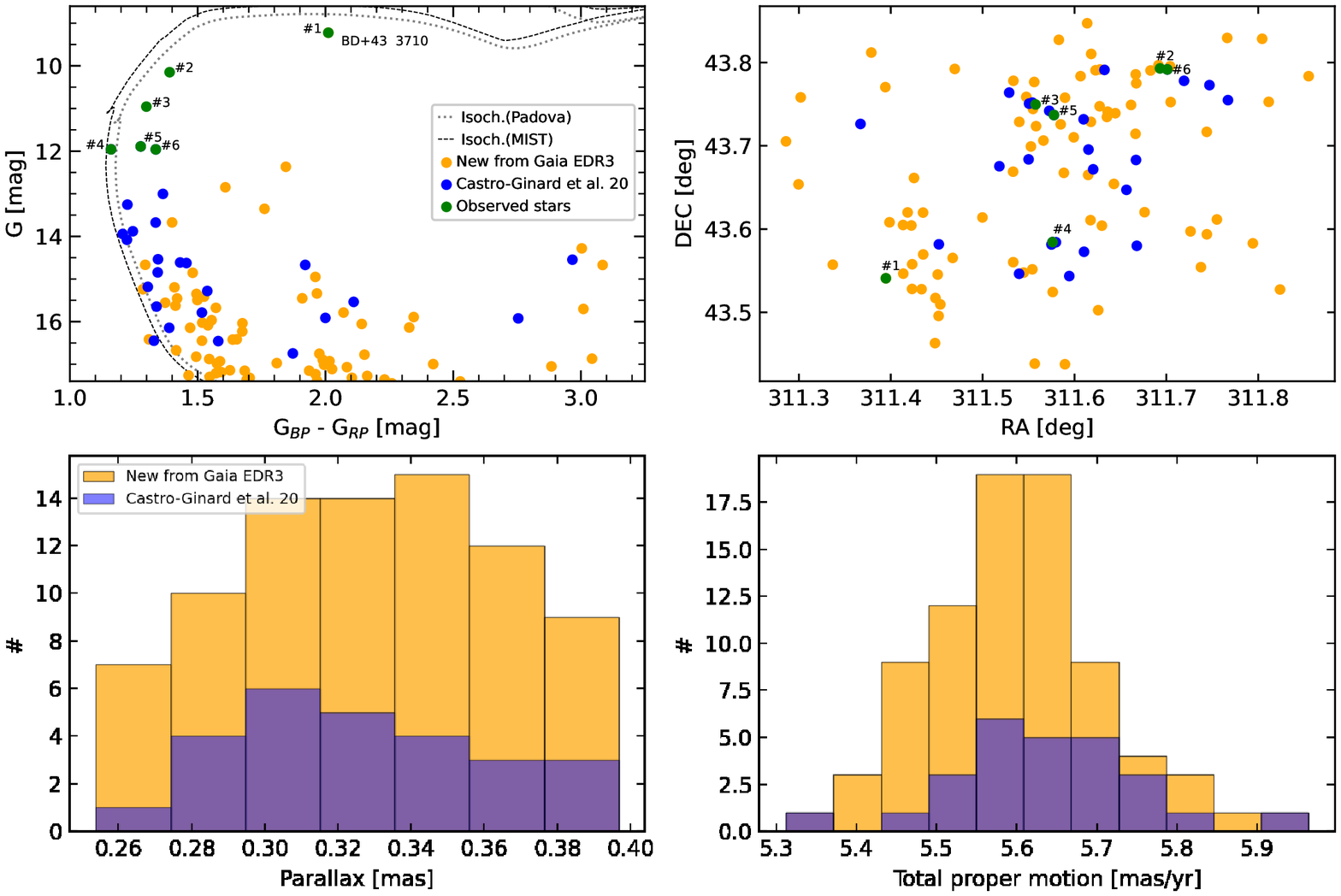}
 \caption{Summary figure for UBC~587. \textit{Top two figures:} Colour-magnitude diagram (left) and sky coordinates of the sources (right). \textit{Bottom two figures:} Histograms of the parallax (left) and total proper motion (right). See Fig.~\ref{figure.ubc103} for a description of the symbols and colours. The grey dashed and dotted lines correspond to isochrones from Padova and MIST databases, respectively, both for a value of 16~Ma and $A_{V}=3.3$.}
 \label{figure.ubc587}
 \end{figure*}
 
 The field lies in the area of the Cygnus X complex ($l=83\fdg6$, $b=+0\fdg3$), less than $2\degr$ away from Deneb, in a gap between two dark clouds. 
 We observed the three brightest members from the list of \citet{castro-ginard20}, two catalogued OB stars in their immediate vicinity, LS~III~$+43\degr$11 (= ALS~11533) and BD~$+43\degr$3714, and a more distant catalogued OB star, BD~$+43\degr$3710, whose proper motions are very similar to those of the cluster. Their spectra are displayed in Fig.~\ref{spectra587}. Their spectral types are presented in Table~\ref{tab:observations}; bottom panel). The two BD stars are early-B supergiants, while the other four are dwarfs with spectral type close to B0.

 Our search for new members in {\em Gaia} EDR3 adds a considerable number of faint members and a few objects with redder colours than the main stellar sequence (see top-left panel of Fig.~\ref{figure.ubc587}). Most of these red objects lie to the West of the field (RA$<311\fdg5$ in the top-right panel of Fig.~\ref{figure.ubc587}), where the dark cloud starts to be noticeable, and thus they are likely more reddened members. According to our criteria, BD~$+43\degr$3710, which is located in the same area, is a cluster member. This object is much brighter than any other member, in good agreement with its spectral classification as a B1\,Ia supergiant. Its red colour indicates that it is much more heavily reddened than the other members with spectra, although it is not redder than other EDR3 members in its vicinity. 
 
 BD~$+43\degr$3710 is surrounded by a complex nebula seen in \textit{Spitzer}/MIPS 24 $\mu$m images \citep[their fig.~1]{kraemer10}. The nebula, which is markedly bipolar, was analysed by \citet{kraemer10} under the erroneous assumption that BD~$+43\degr$3710 was a nearby Carbon star. \citet{flagey14} reported the mistake and used low-resolution near-IR spectra to classify the star as a B-type supergiant. They estimated a spectral type B5 by comparison to a very limited number of standards. To check the possible effect of circumstellar material on its colours, we use its 2MASS photometry, $J=6.60\pm0.02$, $H=6.14\pm0.02$ and $K_{\mathrm{S}}=5.88\pm0.02$. Its $Q_{\mathrm{IR}}= -0.01$ is typical of normal early-type stars, while emission-line stars and other objects with circumstellar discs usually have a more negative $Q_{\mathrm{IR}}$ \citep[e.g.][]{ns07, maiz20}. In addition, the nebula is not detected at wavelengths shorter than 24 $\mu$m and the \textit{Gaia} colours are not redder than those of other members in its vicinity. Therefore, there is no evidence for circumstellar emission in the near-IR. 
 We can thus use the 2MASS photometry to estimate its intrinsic brightness. Assuming a standard \citep{riekes} extinction law, the intrinsic colours of \citet{winkler97} and the $DM=12.4$ found for the cluster, the intrinsic $M_{K}$ is a phenomenal $-7.1$, fully consistent with a Ia luminosity class.

The second supergiant in the field, BD $+43\degr$3714, is not selected as a cluster member. Although its parallax, $\varpi=0.28\pm0.02\:$mas (uncorrected) is fully compatible with the cluster average, its proper motions deviate significantly. For this reason, it is not included in Fig.~\ref{figure.ubc587}. Nevertheless, its observed colour ($BP-RP=1.39$) and $G$ magnitude (1.8~mag brighter than the main sequence stars observed) are fully consistent with a star of its spectral type at the cluster distance. We can thus speculate with a cluster member whose kinematic properties are due to a past binary interaction. In the case of the other catalogued OB star, LS~III~$+43\degr$11, its $\varpi$ ($0.22\pm0.04$~mas) and $\mu_{\delta}$ ($-5.18\pm0.04\:$mas\,a$^{-1}$) are just outside the 2-$\sigma$ threshold used for member selection (but its high RUWE = 2.2 suggests that the errors may be underestimated, even without applying the correction of \citealt{maiz21}, and so it is likely a cluster member). We might speculate that this is a background star, but then it is \textit{far too} bright for its spectral type. Even at the cluster distance, its $M_{\mathrm{K}}=-4.0$ (derived as above) is too bright for the spectral type, although compatible with, for example, a binary.

The age of UBC~587 cannot be strongly constrained beyond stating that it is a very young open cluster. The spectral types of the main sequence stars present force it to be younger than $\sim10\:$Ma, while the presence of a B supergiant, i.e. a moderately evolved massive star, makes it older than $\sim3\:$Ma. The extinction $A_V=3.3$ used in the fit (Fig.~\ref{figure.ubc587}) is adequate for the less reddened members, but much higher values should be used for the more obscured objects.
  
With a \textit{Gaia} distance of 3.0~kpc, UBC~587 is decidedly behind the Cygnus complex. Parallax distances to different star-forming regions in the Cygnus X complex are compatible with a distance around 1.4~kpc \citep{rygl12}. An analysis of \textit{Gaia} DR2 data suggests that the bulk of Cyg~OB2 is considerably farther away, at around 1.8~kpc \citep{berlanas19}, but this is still much closer than UBC~587. Our line of sight in this direction ($l\approx83\degr$) is believed to run along the extent of the Local Arm for several kpc \citep{xu13}. As a very young open cluster, UBC~587 must be a tracer of more distant stretches of this arm than are usually considered (of all the tracers used by \citealt{xu13}, only two have comparable distances). 

\begin{figure}
\centering
\resizebox{\columnwidth}{!}{\includegraphics{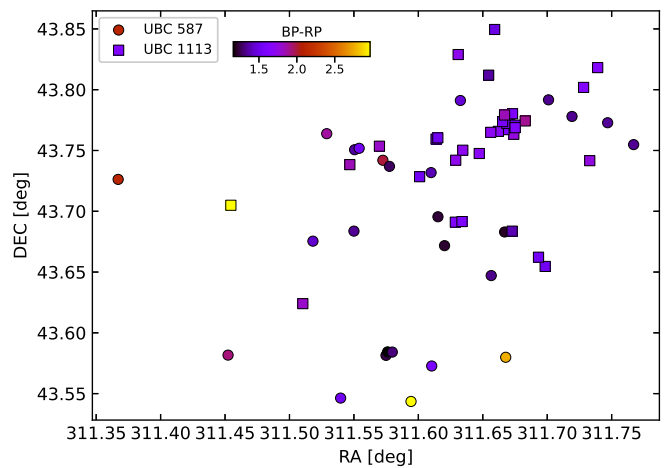}}
\caption{Spatial distribution in sky coordinates of members in UBC~587 and UBC~1113 from \citet{castro-ginard20} (circles) and \citet{castro-ginard22} (with squares), respectively. The colour code corresponds to the $(BP-RP)$ colour. Compare to Fig.~\ref{figure.ubc587}, top-right panel, where a group of EDR3 members with much higher reddening can be seen around $l=311\fdg4$ (close to star \#1).}
\label{fig:ubc587_1113}
\end{figure}

Interestingly, \citet{castro-ginard22} have recently reported a second young open cluster that overlaps on the sky with UBC~587. UBC~1113 has different proper motions, but a very similar parallax, and is affected by about the same amount of extinction. In particular, average values of $(BP-RP)$ for both clusters are very similar, with $(BP-RP)_{UBC~587}$\,=\,1.73$\pm$0.45 and $(BP-RP)_{UBC~1113}$\,=\,1.68$\pm$0.27. Fig.~\ref{fig:ubc587_1113} shows the spatial distribution of members of the two clusters. Although they clearly overlap in space and reddening, UBC~1113 is clearly more spatially concentrated. Inspection of sky images suggests that both clusters are seen through a gap in the dense clouds of the Cygnus~X complex due to nearby star formation.


\section{Conclusions}

We have explored the use of \textit{Gaia} photometry to identify young open clusters among the many new catalogues of candidates generated by means of artificial intelligence algorithms. We find that the presence of an almost vertical sequence in the $(BP-RP)$~vs.~$G$ CMD is a clear-cut signature of a cluster containing B-type stars. The vertical sequence is more readily identifiable when the cluster sequence extends to fainter magnitudes, as in the examples displayed in Fig.~\ref{compare}, but is a reliable signal even when dealing with poor sequences affected by differential reddening, as was the case of UBC~587. Automated algorithms to derive parameters will always give back a cluster age, even in those cases where the data do not provide a definite answer, as in the example of UBC~94 (Fig.~\ref{typical}), and therefore the possibility of confirming the young nature of an open cluster from direct inspection of the CMD is reassuring.

The length of the vertical stripe is directly related to the age of the cluster. In very young open clusters, containing O-type stars and/or blue supergiants, it can extend over more than 6 magnitudes, while Fig.~\ref{compare} illustrates the case of older clusters. The vertical sequence can even be detected in well-populated clusters of 200\,--\,$300\:$Ma. At this age, the main-sequence turn-off should be around B8 at solar metallicity, and the red clump should already be significant, allowing an independent determination of the age. A good example of such a cluster (if somewhat extreme in terms of its mass) is M~11, which is around 220~Ma old \citep{Bav16} and presents a vertical sequence extending for about 2 magnitudes.

Beyond confirming the validity of this approach, two of our test targets have turned out to possess high astrophysical interest of their own. UBC~103 is a moderately massive cluster of around 70 Ma, forming the core of an extended association together with its twin NGC~6683. At a distance around 3~kpc, this region may be associated to the Scutum Arm, although perhaps it represents the result of a high-pitch angle structure, similar to that found by \citet{kuhn21}, as it is aligned with the similarly-aged massive cluster Valparaiso~1, which -- at 2.3~kpc -- seems to lie in between arms. UBC~587 is also a valuable spiral tracer, lying well behind the known structures in the Cygnus complex, and perhaps marking the outer edge of the Local Arm. It contains a very bright B1\,Ia supergiant, BD~$+43\degr$3710, which is surrounded by a complex mid-IR nebula. UBC~184, on the other hand, represents an example of the difficulty in assigning ages to clusters if their stellar sequences are not well populated.

Automated determination of cluster parameters with artificial intelligence algorithms, as used by \citet{cantat20} and \citet{castro-ginard22}, gives results that do not differ very strongly from those obtained by means of visual fits. However, in most cases where a comparison has been done, the value of extinction found by the algorithms has been lower than that found by humans, with no example in our sample of the opposite. As shown by the case of IC~4725 in Fig.~\ref{compare}, the algorithms may be misinterpreting the spread caused by differential reddening. If this is the general case, then extinctions obtained from these artificial intelligence methods are usually underestimated, likely resulting in an overestimation of the cluster age. The sample studied here is too small to claim a systematic effect, but a trend in exactly the same sense can be seen in other clusters that we have recently studied by using \textit{Gaia} photometry, such as NGC~6649 and NGC~6664 \citep{alonso20} or Valparaiso~1 \citep{negueruela21}. This is particularly relevant if we consider that, for example, \citet{dias21}, who use a different isochrone fitting algorithm, find values of extinction that are systematically smaller than those of \citet{cantat20}, except for very high values ($A_V\ga3$), when they become systematically larger.

Intriguingly, all the clusters studied in the direction of the Scutum Star Cloud (UBC~103 and NGC~6683 in this work, and Valparaiso~1 in \citealt{negueruela21}) would give a better isochrone fit for a shorter distance modulus than derived from \textit{Gaia}. We have not seen a similar effect for any other young cluster, with \textit{Gaia} distances generally providing the best isochrone fits. This is an issue that deserves further investigation.


\begin{acknowledgements}

This research is partially supported by the Spanish Government Ministerio de Ciencia e Innovaci\'on and Agencia Estatal de Investigaci\'on (MCIN/AEI/10.130~39/501~100~011~033/FEDER, UE) under grants PGC2018-093741-B-C21/C22 and PID2021-122397NB-C21/C22, and by the Generalitat Valenciana under grants PROMETEO/2019/041 and ASFAE/2022/017.

Based on observations made with the Nordic Optical Telescope, owned
      in collaboration by the University of Turku and Aarhus University,
      and operated jointly by Aarhus University, the University of Turku
      and the University of Oslo, representing Denmark, Finland and
      Norway, the University of Iceland and Stockholm University at the
      Observatorio del Roque de los Muchachos, La Palma, Spain, of the
      Instituto de Astrofisica de Canarias. Data presented here were obtained (Sect.~\ref{subsection.observations}) with ALFOSC, which is provided by the Instituto de Astrofisica de Andaluc\'{\i}a (IAA) under a joint agreement with the University of Copenhagen and the Nordic Optical Telescope Scientific Association (NOTSA).

This work has made use of data from the European Space Agency (ESA) mission
{\it Gaia} (\url{https://www.cosmos.esa.int/gaia}), processed by the {\it Gaia}
Data Processing and Analysis Consortium (DPAC,
\url{https://www.cosmos.esa.int/web/gaia/dpac/consortium}). Funding for the DPAC
has been provided by national institutions, in particular the institutions
participating in the {\it Gaia} Multilateral Agreement.

This research has made use of the Simbad, Vizier and Aladin services developed at the Centre de Donn\'ees Astronomiques de Strasbourg, France. It also makes use of data products from the Two Micron All Sky Survey, which is a joint project of the University of Massachusetts and the Infrared Processing and Analysis
Center/California Institute of Technology, funded by the National
Aeronautics and Space Administration and the National Science
Foundation. We have made extensive use of the TOPCAT \citep{TOPCAT2005} tool. 

\end{acknowledgements}

%
%
\bibliographystyle{aa} 
\bibliography{clusters,gaia,own,cepheids}

\begin{thebibliography}{49}
\expandafter\ifx\csname natexlab\endcsname\relax\def\natexlab#1{#1}\fi

\bibitem[{{Alonso-Santiago} {et~al.}(2020){Alonso-Santiago}, {Negueruela},
  {Marco}, {Tabernero}, \& {Castro}}]{alonso20}
{Alonso-Santiago}, J., {Negueruela}, I., {Marco}, A., {Tabernero}, H.~M., \&
  {Castro}, N. 2020, \aap, 644, A136

\bibitem[{{Anderson} {et~al.}(2016){Anderson}, {Casertano}, {Riess}, {Melis},
  {Holl}, {Semaan}, {Papics}, {Blanco-Cuaresma}, {Eyer}, {Mowlavi},
  {Palaversa}, \& {Roelens}}]{anderson16}
{Anderson}, R.~I., {Casertano}, S., {Riess}, A.~G., {et~al.} 2016, \apjs, 226,
  18

\bibitem[{{Balaguer-N{\'u}{\~n}ez} {et~al.}(2020){Balaguer-N{\'u}{\~n}ez},
  {L{\'o}pez del Fresno}, {Solano}, {Galad{\'\i}-Enr{\'\i}quez}, {Jordi},
  {Jimenez-Esteban}, {Masana}, {Carbajo-Hijarrubia}, \& {Paunzen}}]{balaguer20}
{Balaguer-N{\'u}{\~n}ez}, L., {L{\'o}pez del Fresno}, M., {Solano}, E.,
  {et~al.} 2020, \mnras, 492, 5811

\bibitem[{{Bavarsad} {et~al.}(2016){Bavarsad}, {Sandquist}, {Shetrone}, \&
  {Orosz}}]{Bav16}
{Bavarsad}, E.~A., {Sandquist}, E.~L., {Shetrone}, M.~D., \& {Orosz}, J.~A.
  2016, \apj, 831, 48

\bibitem[{{Berlanas} {et~al.}(2019){Berlanas}, {Wright}, {Herrero}, {Drew}, \&
  {Lennon}}]{berlanas19}
{Berlanas}, S.~R., {Wright}, N.~J., {Herrero}, A., {Drew}, J.~E., \& {Lennon},
  D.~J. 2019, \mnras, 484, 1838

\bibitem[{{Cabrera-Cano} \& {Alfaro}(1990)}]{cabreracano90}
{Cabrera-Cano}, J. \& {Alfaro}, E.~J. 1990, \aap, 235, 94

\bibitem[{{Cantat-Gaudin} {et~al.}(2020){Cantat-Gaudin}, {Anders},
  {Castro-Ginard}, {Jordi}, {Romero-G{\'o}mez}, {Soubiran}, {Casamiquela},
  {Tarricq}, {Moitinho}, {Vallenari}, {Bragaglia}, {Krone-Martins}, \&
  {Kounkel}}]{cantat20}
{Cantat-Gaudin}, T., {Anders}, F., {Castro-Ginard}, A., {et~al.} 2020, \aap,
  640, A1

\bibitem[{{Cantat-Gaudin} {et~al.}(2018){Cantat-Gaudin}, {Jordi}, {Vallenari},
  {Bragaglia}, {Balaguer-N{\'u}{\~n}ez}, {Soubiran}, {Bossini}, {Moitinho},
  {Castro-Ginard}, {Krone-Martins}, {Casamiquela}, {Sordo}, \&
  {Carrera}}]{cantat18}
{Cantat-Gaudin}, T., {Jordi}, C., {Vallenari}, A., {et~al.} 2018, \aap, 618,
  A93

\bibitem[{{Cardelli} {et~al.}(1989){Cardelli}, {Clayton}, \&
  {Mathis}}]{cardelli}
{Cardelli}, J.~A., {Clayton}, G.~C., \& {Mathis}, J.~S. 1989, \apj, 345, 245

\bibitem[{{Castro-Ginard} {et~al.}(2020){Castro-Ginard}, {Jordi}, {Luri},
  {{\'A}lvarez Cid-Fuentes}, {Casamiquela}, {Anders}, {Cantat-Gaudin},
  {Mongui{\'o}}, {Balaguer-N{\'u}{\~n}ez}, {Sol{\`a}}, \&
  {Badia}}]{castro-ginard20}
{Castro-Ginard}, A., {Jordi}, C., {Luri}, X., {et~al.} 2020, \aap, 635, A45

\bibitem[{{Castro-Ginard} {et~al.}(2019){Castro-Ginard}, {Jordi}, {Luri},
  {Cantat-Gaudin}, \& {Balaguer-N{\'u}{\~n}ez}}]{castro-ginard19}
{Castro-Ginard}, A., {Jordi}, C., {Luri}, X., {Cantat-Gaudin}, T., \&
  {Balaguer-N{\'u}{\~n}ez}, L. 2019, \aap, 627, A35

\bibitem[{{Castro-Ginard} {et~al.}(2022){Castro-Ginard}, {Jordi}, {Luri},
  {Cantat-Gaudin}, {Carrasco}, {Casamiquela}, {Anders},
  {Balaguer-N{\'u}{\~n}ez}, \& {Badia}}]{castro-ginard22}
{Castro-Ginard}, A., {Jordi}, C., {Luri}, X., {et~al.} 2022, \aap, 661, A118

\bibitem[{{Castro-Ginard} {et~al.}(2021){Castro-Ginard}, {McMillan}, {Luri},
  {Jordi}, {Romero-G{\'o}mez}, {Cantat-Gaudin}, {Casamiquela}, {Tarricq},
  {Soubiran}, \& {Anders}}]{castro-ginard21}
{Castro-Ginard}, A., {McMillan}, P.~J., {Luri}, X., {et~al.} 2021, \aap, 652,
  A162

\bibitem[{{Chen} {et~al.}(2015){Chen}, {Bressan}, {Girardi}, {Marigo}, {Kong},
  \& {Lanza}}]{chen15}
{Chen}, Y., {Bressan}, A., {Girardi}, L., {et~al.} 2015, \mnras, 452, 1068

\bibitem[{{Dias} {et~al.}(2021){Dias}, {Monteiro}, {Moitinho}, {L{\'e}pine},
  {Carraro}, {Paunzen}, {Alessi}, \& {Villela}}]{dias21}
{Dias}, W.~S., {Monteiro}, H., {Moitinho}, A., {et~al.} 2021, \mnras, 504, 356

\bibitem[{{Ekstr{\"o}m} {et~al.}(2012){Ekstr{\"o}m}, {Georgy}, {Eggenberger},
  {Meynet}, {Mowlavi}, {Wyttenbach}, {Granada}, {Decressin}, {Hirschi},
  {Frischknecht}, {Charbonnel}, \& {Maeder}}]{ekstrom12}
{Ekstr{\"o}m}, S., {Georgy}, C., {Eggenberger}, P., {et~al.} 2012, \aap, 537,
  A146

\bibitem[{{Ferreira} {et~al.}(2020){Ferreira}, {Corradi}, {Maia}, {Angelo}, \&
  {Santos}}]{ferreira20}
{Ferreira}, F.~A., {Corradi}, W.~J.~B., {Maia}, F.~F.~S., {Angelo}, M.~S., \&
  {Santos}, J.~F.~C., J. 2020, \mnras, 496, 2021

\bibitem[{{Flagey} {et~al.}(2014){Flagey}, {Noriega-Crespo}, {Petric}, \&
  {Geballe}}]{flagey14}
{Flagey}, N., {Noriega-Crespo}, A., {Petric}, A., \& {Geballe}, T.~R. 2014,
  \aj, 148, 34

\bibitem[{{Gaia Collaboration} {et~al.}(2018{\natexlab{a}}){Gaia
  Collaboration}, {Brown}, {Vallenari}, {Prusti}, {de Bruijne}, {Babusiaux},
  {Bailer-Jones}, {Biermann}, {Evans}, {Eyer}, \& et~al.}]{brown18}
{Gaia Collaboration}, {Brown}, A.~G.~A., {Vallenari}, A., {et~al.}
  2018{\natexlab{a}}, \aap, 616, A1

\bibitem[{{Gaia Collaboration} {et~al.}(2018{\natexlab{b}}){Gaia
  Collaboration}, {Brown}, {Vallenari}, {Prusti}, {de Bruijne}, {Babusiaux},
  {Bailer-Jones}, {Biermann}, {Evans}, {Eyer}, {Jansen}, {Jordi}, {Klioner},
  {Lammers}, {Lindegren}, {Luri}, {Mignard}, {Panem}, {Pourbaix}, {Randich},
  {Sartoretti}, {Siddiqui}, {Soubiran}, {van Leeuwen}, {Walton}, {Arenou},
  {Bastian}, {Cropper}, {Drimmel}, {Katz}, {Lattanzi}, {Bakker}, {Cacciari},
  {Casta{\~n}eda}, {Chaoul}, {Cheek}, {De Angeli}, {Fabricius}, {Guerra},
  {Holl}, {Masana}, {Messineo}, {Mowlavi}, {Nienartowicz}, {Panuzzo},
  {Portell}, {Riello}, {Seabroke}, {Tanga}, {Th{\'e}venin}, {Gracia-Abril},
  {Comoretto}, {Garcia-Reinaldos}, {Teyssier}, {Altmann}, {Andrae}, {Audard},
  {Bellas-Velidis}, {Benson}, {Berthier}, {Blomme}, {Burgess}, {Busso},
  {Carry}, {Cellino}, {Clementini}, {Clotet}, {Creevey}, {Davidson}, {De
  Ridder}, {Delchambre}, {Dell'Oro}, {Ducourant},
  {Fern{\'a}ndez-Hern{\'a}ndez}, {Fouesneau}, {Fr{\'e}mat}, {Galluccio},
  {Garc{\'\i}a-Torres}, {Gonz{\'a}lez-N{\'u}{\~n}ez}, {Gonz{\'a}lez-Vidal},
  {Gosset}, {Guy}, {Halbwachs}, {Hambly}, {Harrison}, {Hern{\'a}ndez},
  {Hestroffer}, {Hodgkin}, {Hutton}, {Jasniewicz}, {Jean-Antoine-Piccolo},
  {Jordan}, {Korn}, {Krone-Martins}, {Lanzafame}, {Lebzelter}, {L{\"o}ffler},
  {Manteiga}, {Marrese}, {Mart{\'\i}n-Fleitas}, {Moitinho}, {Mora}, {Muinonen},
  {Osinde}, {Pancino}, {Pauwels}, {Petit}, {Recio-Blanco}, {Richards},
  {Rimoldini}, {Robin}, {Sarro}, {Siopis}, {Smith}, {Sozzetti}, {S{\"u}veges},
  {Torra}, {van Reeven}, {Abbas}, {Abreu Aramburu}, {Accart}, {Aerts},
  {Altavilla}, {{\'A}lvarez}, {Alvarez}, {Alves}, {Anderson}, {Andrei},
  {Anglada Varela}, {Antiche}, {Antoja}, {Arcay}, {Astraatmadja}, {Bach},
  {Baker}, {Balaguer-N{\'u}{\~n}ez}, {Balm}, {Barache}, {Barata}, {Barbato},
  {Barblan}, {Barklem}, {Barrado}, {Barros}, {Barstow}, {Bartholom{\'e}
  Mu{\~n}oz}, {Bassilana}, {Becciani}, {Bellazzini}, {Berihuete}, {Bertone},
  {Bianchi}, {Bienaym{\'e}}, {Blanco-Cuaresma}, {Boch}, {Boeche}, {Bombrun},
  {Borrachero}, {Bossini}, {Bouquillon}, {Bourda}, {Bragaglia}, {Bramante},
  {Breddels}, {Bressan}, {Brouillet}, {Br{\"u}semeister}, {Brugaletta},
  {Bucciarelli}, {Burlacu}, {Busonero}, {Butkevich}, {Buzzi}, {Caffau},
  {Cancelliere}, {Cannizzaro}, {Cantat-Gaudin}, {Carballo}, {Carlucci},
  {Carrasco}, {Casamiquela}, {Castellani}, {Castro-Ginard}, {Charlot},
  {Chemin}, {Chiavassa}, {Cocozza}, {Costigan}, {Cowell}, {Crifo}, {Crosta},
  {Crowley}, {Cuypers}, {Dafonte}, {Damerdji}, {Dapergolas}, {David}, {David},
  {de Laverny}, {De Luise}, {De March}, {de Martino}, {de Souza}, {de Torres},
  {Debosscher}, {del Pozo}, {Delbo}, {Delgado}, {Delgado}, {Di Matteo},
  {Diakite}, {Diener}, {Distefano}, {Dolding}, {Drazinos}, {Dur{\'a}n},
  {Edvardsson}, {Enke}, {Eriksson}, {Esquej}, {Eynard Bontemps}, {Fabre},
  {Fabrizio}, {Faigler}, {Falc{\~a}o}, {Farr{\`a}s Casas}, {Federici},
  {Fedorets}, {Fernique}, {Figueras}, {Filippi}, {Findeisen}, {Fonti},
  {Fraile}, {Fraser}, {Fr{\'e}zouls}, {Gai}, {Galleti}, {Garabato},
  {Garc{\'\i}a-Sedano}, {Garofalo}, {Garralda}, {Gavel}, {Gavras}, {Gerssen},
  {Geyer}, {Giacobbe}, {Gilmore}, {Girona}, {Giuffrida}, {Glass}, {Gomes},
  {Granvik}, {Gueguen}, {Guerrier}, {Guiraud}, {Guti{\'e}rrez-S{\'a}nchez},
  {Haigron}, {Hatzidimitriou}, {Hauser}, {Haywood}, {Heiter}, {Helmi}, {Heu},
  {Hilger}, {Hobbs}, {Hofmann}, {Holland}, {Huckle}, {Hypki}, {Icardi},
  {Jan{\ss}en}, {Jevardat de Fombelle}, {Jonker}, {Juh{\'a}sz}, {Julbe},
  {Karampelas}, {Kewley}, {Klar}, {Kochoska}, {Kohley}, {Kolenberg},
  {Kontizas}, {Kontizas}, {Koposov}, {Kordopatis}, {Kostrzewa-Rutkowska},
  {Koubsky}, {Lambert}, {Lanza}, {Lasne}, {Lavigne}, {Le Fustec}, {Le
  Poncin-Lafitte}, {Lebreton}, {Leccia}, {Leclerc}, {Lecoeur-Taibi},
  {Lenhardt}, {Leroux}, {Liao}, {Licata}, {Lindstr{\o}m}, {Lister}, {Livanou},
  {Lobel}, {L{\'o}pez}, {Managau}, {Mann}, {Mantelet}, {Marchal}, {Marchant},
  {Marconi}, {Marinoni}, {Marschalk{\'o}}, {Marshall}, {Martino}, {Marton},
  {Mary}, {Massari}, {Matijevi{\v{c}}}, {Mazeh}, {McMillan}, {Messina},
  {Michalik}, {Millar}, {Molina}, {Molinaro}, {Moln{\'a}r}, {Montegriffo},
  {Mor}, {Morbidelli}, {Morel}, {Morris}, {Mulone}, {Muraveva}, {Musella},
  {Nelemans}, {Nicastro}, {Noval}, {O'Mullane}, {Ord{\'e}novic},
  {Ord{\'o}{\~n}ez-Blanco}, {Osborne}, {Pagani}, {Pagano}, {Pailler},
  {Palacin}, {Palaversa}, {Panahi}, {Pawlak}, {Piersimoni}, {Pineau}, {Plachy},
  {Plum}, {Poggio}, {Poujoulet}, {Pr{\v{s}}a}, {Pulone}, {Racero}, {Ragaini},
  {Rambaux}, {Ramos-Lerate}, {Regibo}, {Reyl{\'e}}, {Riclet}, {Ripepi}, {Riva},
  {Rivard}, {Rixon}, {Roegiers}, {Roelens}, {Romero-G{\'o}mez}, {Rowell},
  {Royer}, {Ruiz-Dern}, {Sadowski}, {Sagrist{\`a} Sell{\'e}s}, {Sahlmann},
  {Salgado}, {Salguero}, {Sanna}, {Santana-Ros}, {Sarasso}, {Savietto},
  {Schultheis}, {Sciacca}, {Segol}, {Segovia}, {S{\'e}gransan}, {Shih},
  {Siltala}, {Silva}, {Smart}, {Smith}, {Solano}, {Solitro}, {Sordo}, {Soria
  Nieto}, {Souchay}, {Spagna}, {Spoto}, {Stampa}, {Steele},
  {Steidelm{\"u}ller}, {Stephenson}, {Stoev}, {Suess}, {Surdej}, {Szabados},
  {Szegedi-Elek}, {Tapiador}, {Taris}, {Tauran}, {Taylor}, {Teixeira},
  {Terrett}, {Teyssand ier}, {Thuillot}, {Titarenko}, {Torra Clotet}, {Turon},
  {Ulla}, {Utrilla}, {Uzzi}, {Vaillant}, {Valentini}, {Valette}, {van Elteren},
  {Van Hemelryck}, {van Leeuwen}, {Vaschetto}, {Vecchiato}, {Veljanoski},
  {Viala}, {Vicente}, {Vogt}, {von Essen}, {Voss}, {Votruba}, {Voutsinas},
  {Walmsley}, {Weiler}, {Wertz}, {Wevers}, {Wyrzykowski}, {Yoldas},
  {{\v{Z}}erjal}, {Ziaeepour}, {Zorec}, {Zschocke}, {Zucker}, {Zurbach}, \&
  {Zwitter}}]{2018A&A...616A...1G}
{Gaia Collaboration}, {Brown}, A.~G.~A., {Vallenari}, A., {et~al.}
  2018{\natexlab{b}}, \aap, 616, A1

\bibitem[{{Gaia Collaboration} {et~al.}(2021){Gaia Collaboration}, {Brown},
  {Vallenari}, {Prusti}, {de Bruijne}, {Babusiaux}, {Biermann}, {Creevey},
  {Evans}, {Eyer}, {Hutton}, {Jansen}, {Jordi}, {Klioner}, {Lammers},
  {Lindegren}, {Luri}, {Mignard}, {Panem}, {Pourbaix}, {Randich}, {Sartoretti},
  {Soubiran}, {Walton}, {Arenou}, {Bailer-Jones}, {Bastian}, {Cropper},
  {Drimmel}, {Katz}, {Lattanzi}, {van Leeuwen}, {Bakker}, {Cacciari},
  {Casta{\~n}eda}, {De Angeli}, {Ducourant}, {Fabricius}, {Fouesneau},
  {Fr{\'e}mat}, {Guerra}, {Guerrier}, {Guiraud}, {Jean-Antoine Piccolo},
  {Masana}, {Messineo}, {Mowlavi}, {Nicolas}, {Nienartowicz}, {Pailler},
  {Panuzzo}, {Riclet}, {Roux}, {Seabroke}, {Sordo}, {Tanga}, {Th{\'e}venin},
  {Gracia-Abril}, {Portell}, {Teyssier}, {Altmann}, {Andrae}, {Bellas-Velidis},
  {Benson}, {Berthier}, {Blomme}, {Brugaletta}, {Burgess}, {Busso}, {Carry},
  {Cellino}, {Cheek}, {Clementini}, {Damerdji}, {Davidson}, {Delchambre},
  {Dell'Oro}, {Fern{\'a}ndez-Hern{\'a}ndez}, {Galluccio}, {Garc{\'\i}a-Lario},
  {Garcia-Reinaldos}, {Gonz{\'a}lez-N{\'u}{\~n}ez}, {Gosset}, {Haigron},
  {Halbwachs}, {Hambly}, {Harrison}, {Hatzidimitriou}, {Heiter},
  {Hern{\'a}ndez}, {Hestroffer}, {Hodgkin}, {Holl}, {Jan{\ss}en}, {Jevardat de
  Fombelle}, {Jordan}, {Krone-Martins}, {Lanzafame}, {L{\"o}ffler}, {Lorca},
  {Manteiga}, {Marchal}, {Marrese}, {Moitinho}, {Mora}, {Muinonen}, {Osborne},
  {Pancino}, {Pauwels}, {Petit}, {Recio-Blanco}, {Richards}, {Riello},
  {Rimoldini}, {Robin}, {Roegiers}, {Rybizki}, {Sarro}, {Siopis}, {Smith},
  {Sozzetti}, {Ulla}, {Utrilla}, {van Leeuwen}, {van Reeven}, {Abbas}, {Abreu
  Aramburu}, {Accart}, {Aerts}, {Aguado}, {Ajaj}, {Altavilla}, {{\'A}lvarez},
  {{\'A}lvarez Cid-Fuentes}, {Alves}, {Anderson}, {Anglada Varela}, {Antoja},
  {Audard}, {Baines}, {Baker}, {Balaguer-N{\'u}{\~n}ez}, {Balbinot}, {Balog},
  {Barache}, {Barbato}, {Barros}, {Barstow}, {Bartolom{\'e}}, {Bassilana},
  {Bauchet}, {Baudesson-Stella}, {Becciani}, {Bellazzini}, {Bernet}, {Bertone},
  {Bianchi}, {Blanco-Cuaresma}, {Boch}, {Bombrun}, {Bossini}, {Bouquillon},
  {Bragaglia}, {Bramante}, {Breedt}, {Bressan}, {Brouillet}, {Bucciarelli},
  {Burlacu}, {Busonero}, {Butkevich}, {Buzzi}, {Caffau}, {Cancelliere},
  {C{\'a}novas}, {Cantat-Gaudin}, {Carballo}, {Carlucci}, {Carnerero},
  {Carrasco}, {Casamiquela}, {Castellani}, {Castro-Ginard}, {Castro Sampol},
  {Chaoul}, {Charlot}, {Chemin}, {Chiavassa}, {Cioni}, {Comoretto}, {Cooper},
  {Cornez}, {Cowell}, {Crifo}, {Crosta}, {Crowley}, {Dafonte}, {Dapergolas},
  {David}, {David}, {de Laverny}, {De Luise}, {De March}, {De Ridder}, {de
  Souza}, {de Teodoro}, {de Torres}, {del Peloso}, {del Pozo}, {Delbo},
  {Delgado}, {Delgado}, {Delisle}, {Di Matteo}, {Diakite}, {Diener},
  {Distefano}, {Dolding}, {Eappachen}, {Edvardsson}, {Enke}, {Esquej}, {Fabre},
  {Fabrizio}, {Faigler}, {Fedorets}, {Fernique}, {Fienga}, {Figueras},
  {Fouron}, {Fragkoudi}, {Fraile}, {Franke}, {Gai}, {Garabato},
  {Garcia-Gutierrez}, {Garc{\'\i}a-Torres}, {Garofalo}, {Gavras}, {Gerlach},
  {Geyer}, {Giacobbe}, {Gilmore}, {Girona}, {Giuffrida}, {Gomel}, {Gomez},
  {Gonzalez-Santamaria}, {Gonz{\'a}lez-Vidal}, {Granvik},
  {Guti{\'e}rrez-S{\'a}nchez}, {Guy}, {Hauser}, {Haywood}, {Helmi}, {Hidalgo},
  {Hilger}, {H{\l}adczuk}, {Hobbs}, {Holland}, {Huckle}, {Jasniewicz},
  {Jonker}, {Juaristi Campillo}, {Julbe}, {Karbevska}, {Kervella}, {Khanna},
  {Kochoska}, {Kontizas}, {Kordopatis}, {Korn}, {Kostrzewa-Rutkowska},
  {Kruszy{\'n}ska}, {Lambert}, {Lanza}, {Lasne}, {Le Campion}, {Le Fustec},
  {Lebreton}, {Lebzelter}, {Leccia}, {Leclerc}, {Lecoeur-Taibi}, {Liao},
  {Licata}, {Lindstr{\o}m}, {Lister}, {Livanou}, {Lobel}, {Madrero Pardo},
  {Managau}, {Mann}, {Marchant}, {Marconi}, {Marcos Santos}, {Marinoni},
  {Marocco}, {Marshall}, {Martin Polo}, {Mart{\'\i}n-Fleitas}, {Masip},
  {Massari}, {Mastrobuono-Battisti}, {Mazeh}, {McMillan}, {Messina},
  {Michalik}, {Millar}, {Mints}, {Molina}, {Molinaro}, {Moln{\'a}r},
  {Montegriffo}, {Mor}, {Morbidelli}, {Morel}, {Morris}, {Mulone}, {Munoz},
  {Muraveva}, {Murphy}, {Musella}, {Noval}, {Ord{\'e}novic}, {Orr{\`u}},
  {Osinde}, {Pagani}, {Pagano}, {Palaversa}, {Palicio}, {Panahi}, {Pawlak},
  {Pe{\~n}alosa Esteller}, {Penttil{\"a}}, {Piersimoni}, {Pineau}, {Plachy},
  {Plum}, {Poggio}, {Poretti}, {Poujoulet}, {Pr{\v{s}}a}, {Pulone}, {Racero},
  {Ragaini}, {Rainer}, {Raiteri}, {Rambaux}, {Ramos}, {Ramos-Lerate}, {Re
  Fiorentin}, {Regibo}, {Reyl{\'e}}, {Ripepi}, {Riva}, {Rixon}, {Robichon},
  {Robin}, {Roelens}, {Rohrbasser}, {Romero-G{\'o}mez}, {Rowell}, {Royer},
  {Rybicki}, {Sadowski}, {Sagrist{\`a} Sell{\'e}s}, {Sahlmann}, {Salgado},
  {Salguero}, {Samaras}, {Sanchez Gimenez}, {Sanna}, {Santove{\~n}a},
  {Sarasso}, {Schultheis}, {Sciacca}, {Segol}, {Segovia}, {S{\'e}gransan},
  {Semeux}, {Shahaf}, {Siddiqui}, {Siebert}, {Siltala}, {Slezak}, {Smart},
  {Solano}, {Solitro}, {Souami}, {Souchay}, {Spagna}, {Spoto}, {Steele},
  {Steidelm{\"u}ller}, {Stephenson}, {S{\"u}veges}, {Szabados}, {Szegedi-Elek},
  {Taris}, {Tauran}, {Taylor}, {Teixeira}, {Thuillot}, {Tonello}, {Torra},
  {Torra}, {Turon}, {Unger}, {Vaillant}, {van Dillen}, {Vanel}, {Vecchiato},
  {Viala}, {Vicente}, {Voutsinas}, {Weiler}, {Wevers}, {Wyrzykowski}, {Yoldas},
  {Yvard}, {Zhao}, {Zorec}, {Zucker}, {Zurbach}, \&
  {Zwitter}}]{2021A&A...649A...1G}
{Gaia Collaboration}, {Brown}, A.~G.~A., {Vallenari}, A., {et~al.} 2021, \aap,
  649, A1

\bibitem[{{Galadi-Enriquez} {et~al.}(1998){Galadi-Enriquez}, {Jordi}, \&
  {Trullols}}]{galadi98}
{Galadi-Enriquez}, D., {Jordi}, C., \& {Trullols}, E. 1998, \aap, 337, 125

\bibitem[{{Hao} {et~al.}(2020){Hao}, {Xu}, {Wu}, {He}, \& {Bian}}]{hao20}
{Hao}, C., {Xu}, Y., {Wu}, Z., {He}, Z., \& {Bian}, S. 2020, \pasp, 132, 034502

\bibitem[{{Hao} {et~al.}(2022){Hao}, {Xu}, {Wu}, {Lin}, {Liu}, \& {Li}}]{hao22}
{Hao}, C.~J., {Xu}, Y., {Wu}, Z.~Y., {et~al.} 2022, \aap, 660, A4

\bibitem[{{Harmanec}(1988)}]{harmanec88}
{Harmanec}, P. 1988, Bulletin of the Astronomical Institutes of Czechoslovakia,
  39, 329

\bibitem[{{He} {et~al.}(2022){He}, {Liu}, {Luo}, {Wang}, \& {Jiang}}]{he22}
{He}, Z., {Liu}, X., {Luo}, Y., {Wang}, K., \& {Jiang}, Q. 2022, arXiv
  e-prints, arXiv:2209.08504

\bibitem[{{He} {et~al.}(2020){He}, {Xu}, {Hao}, {Wu}, \& {Li}}]{he20}
{He}, Z.-H., {Xu}, Y., {Hao}, C.-J., {Wu}, Z.-Y., \& {Li}, J.-J. 2020, arXiv
  e-prints, arXiv:2010.14870

\bibitem[{{Kharchenko} {et~al.}(2013){Kharchenko}, {Piskunov}, {Schilbach},
  {R{\"o}ser}, \& {Scholz}}]{kharchenko13}
{Kharchenko}, N.~V., {Piskunov}, A.~E., {Schilbach}, E., {R{\"o}ser}, S., \&
  {Scholz}, R.-D. 2013, \aap, 558, A53

\bibitem[{{Kraemer} {et~al.}(2010){Kraemer}, {Hora}, {Egan}, {Adams}, {Allen},
  {Bontemps}, {Carey}, {Fazio}, {Gutermuth}, {Keto}, {Koenig}, {Megeath},
  {Mizuno}, {Motte}, {Price}, {Schneider}, {Simon}, \& {Smith}}]{kraemer10}
{Kraemer}, K.~E., {Hora}, J.~L., {Egan}, M.~P., {et~al.} 2010, \aj, 139, 2319

\bibitem[{{Kuhn} {et~al.}(2021){Kuhn}, {Benjamin}, {Zucker}, {Krone-Martins},
  {de Souza}, {Castro-Ginard}, {Ishida}, {Povich}, \& {Hillenbrand}}]{kuhn21}
{Kuhn}, M.~A., {Benjamin}, R.~A., {Zucker}, C., {et~al.} 2021, \aap, 651, L10

\bibitem[{{Lindegren} {et~al.}(2021){Lindegren}, {Bastian}, {Biermann},
  {Bombrun}, {de Torres}, {Gerlach}, {Geyer}, {Hern{\'a}ndez}, {Hilger},
  {Hobbs}, {Klioner}, {Lammers}, {McMillan}, {Ramos-Lerate},
  {Steidelm{\"u}ller}, {Stephenson}, \& {van Leeuwen}}]{2021A&A...649A...4L}
{Lindegren}, L., {Bastian}, U., {Biermann}, M., {et~al.} 2021, \aap, 649, A4

\bibitem[{{Lindegren} {et~al.}(2018){Lindegren}, {Hern{\'a}ndez}, {Bombrun},
  {Klioner}, {Bastian}, {Ramos-Lerate}, {de Torres}, {Steidelm{\"u}ller},
  {Stephenson}, {Hobbs}, {Lammers}, {Biermann}, {Geyer}, {Hilger}, {Michalik},
  {Stampa}, {McMillan}, {Casta{\~n}eda}, {Clotet}, {Comoretto}, {Davidson},
  {Fabricius}, {Gracia}, {Hambly}, {Hutton}, {Mora}, {Portell}, {van Leeuwen},
  {Abbas}, {Abreu}, {Altmann}, {Andrei}, {Anglada}, {Balaguer-N{\'u}{\~n}ez},
  {Barache}, {Becciani}, {Bertone}, {Bianchi}, {Bouquillon}, {Bourda},
  {Br{\"u}semeister}, {Bucciarelli}, {Busonero}, {Buzzi}, {Cancelliere},
  {Carlucci}, {Charlot}, {Cheek}, {Crosta}, {Crowley}, {de Bruijne}, {de
  Felice}, {Drimmel}, {Esquej}, {Fienga}, {Fraile}, {Gai}, {Garralda},
  {Gonz{\'a}lez-Vidal}, {Guerra}, {Hauser}, {Hofmann}, {Holl}, {Jordan},
  {Lattanzi}, {Lenhardt}, {Liao}, {Licata}, {Lister}, {L{\"o}ffler},
  {Marchant}, {Martin-Fleitas}, {Messineo}, {Mignard}, {Morbidelli}, {Poggio},
  {Riva}, {Rowell}, {Salguero}, {Sarasso}, {Sciacca}, {Siddiqui}, {Smart},
  {Spagna}, {Steele}, {Taris}, {Torra}, {van Elteren}, {van Reeven}, \&
  {Vecchiato}}]{2018A&A...616A...2L}
{Lindegren}, L., {Hern{\'a}ndez}, J., {Bombrun}, A., {et~al.} 2018, \aap, 616,
  A2

\bibitem[{{Liu} \& {Pang}(2019)}]{liupang19}
{Liu}, L. \& {Pang}, X. 2019, \apjs, 245, 32

\bibitem[{{Ma{\'\i}z Apell{\'a}niz} {et~al.}(2020{\natexlab{a}}){Ma{\'\i}z
  Apell{\'a}niz}, {Crespo Bellido}, {Barb{\'a}}, {Fern{\'a}ndez Aranda}, \&
  {Sota}}]{maiz_villa1}
{Ma{\'\i}z Apell{\'a}niz}, J., {Crespo Bellido}, P., {Barb{\'a}}, R.~H.,
  {Fern{\'a}ndez Aranda}, R., \& {Sota}, A. 2020{\natexlab{a}}, \aap, 643, A138

\bibitem[{{Ma{\'\i}z Apell{\'a}niz} {et~al.}(2021){Ma{\'\i}z Apell{\'a}niz},
  {Pantaleoni Gonz{\'a}lez}, \& {Barb{\'a}}}]{maiz21}
{Ma{\'\i}z Apell{\'a}niz}, J., {Pantaleoni Gonz{\'a}lez}, M., \& {Barb{\'a}},
  R.~H. 2021, \aap, 649, A13

\bibitem[{{Ma{\'\i}z Apell{\'a}niz} {et~al.}(2020{\natexlab{b}}){Ma{\'\i}z
  Apell{\'a}niz}, {Pantaleoni Gonz{\'a}lez}, {Barb{\'a}}, {Garc{\'\i}a-Lario},
  \& {Nogueras-Lara}}]{maiz20}
{Ma{\'\i}z Apell{\'a}niz}, J., {Pantaleoni Gonz{\'a}lez}, M., {Barb{\'a}},
  R.~H., {Garc{\'\i}a-Lario}, P., \& {Nogueras-Lara}, F. 2020{\natexlab{b}},
  \mnras, 496, 4951

\bibitem[{{Ma{\'\i}z Apell{\'a}niz} \& {Weiler}(2018)}]{mapw18}
{Ma{\'\i}z Apell{\'a}niz}, J. \& {Weiler}, M. 2018, \aap, 619, A180

\bibitem[{{Negueruela}(2017)}]{negueruela17}
{Negueruela}, I. 2017, in The Lives and Death-Throes of Massive Stars, ed.
  J.~J. {Eldridge}, J.~C. {Bray}, L.~A.~S. {McClelland}, \& L.~{Xiao}, Vol.
  329, 271--278

\bibitem[{{Negueruela} {et~al.}(2021){Negueruela}, {Chen{\'e}}, {Tabernero},
  {Dorda}, {Borissova}, {Marco}, \& {Kurtev}}]{negueruela21}
{Negueruela}, I., {Chen{\'e}}, A.~N., {Tabernero}, H.~M., {et~al.} 2021,
  \mnras, 505, 1618

\bibitem[{{Negueruela} {et~al.}(2018){Negueruela}, {Mongui{\'o}}, {Marco},
  {Tabernero}, {Gonz{\'a}lez-Fern{\'a}ndez}, \& {Dorda}}]{negueruela18}
{Negueruela}, I., {Mongui{\'o}}, M., {Marco}, A., {et~al.} 2018, \mnras, 477,
  2976

\bibitem[{{Negueruela} \& {Schurch}(2007)}]{ns07}
{Negueruela}, I. \& {Schurch}, M.~P.~E. 2007, \aap, 461, 631

\bibitem[{{Rieke} \& {Lebofsky}(1985)}]{riekes}
{Rieke}, G.~H. \& {Lebofsky}, M.~J. 1985, \apj, 288, 618

\bibitem[{{Roslund}(1963)}]{roslund63}
{Roslund}, C. 1963, Arkiv for Astronomi, 3, 97

\bibitem[{{Rygl} {et~al.}(2012){Rygl}, {Brunthaler}, {Sanna}, {Menten}, {Reid},
  {van Langevelde}, {Honma}, {Torstensson}, \& {Fujisawa}}]{rygl12}
{Rygl}, K.~L.~J., {Brunthaler}, A., {Sanna}, A., {et~al.} 2012, \aap, 539, A79

\bibitem[{{Sanders}(1971)}]{sanders71}
{Sanders}, W.~L. 1971, \aap, 14, 226

\bibitem[{{Taylor}(2005)}]{TOPCAT2005}
{Taylor}, M.~B. 2005, in Astronomical Society of the Pacific Conference Series,
  Vol. 347, Astronomical Data Analysis Software and Systems XIV, ed.
  P.~{Shopbell}, M.~{Britton}, \& R.~{Ebert}, 29

\bibitem[{{Winkler}(1997)}]{winkler97}
{Winkler}, H. 1997, \mnras, 287, 481

\bibitem[{{Xu} {et~al.}(2013){Xu}, {Li}, {Reid}, {Menten}, {Zheng},
  {Brunthaler}, {Moscadelli}, {Dame}, \& {Zhang}}]{xu13}
{Xu}, Y., {Li}, J.~J., {Reid}, M.~J., {et~al.} 2013, \apj, 769, 15

\bibitem[{{Yilmaz}(1966)}]{yilmaz66}
{Yilmaz}, F. 1966, \zap, 64, 61

\end{thebibliography}

\begin{appendix} 

\section{Parameters of NGC~6683}
\label{app:ngc}

As discussed in Section~\ref{subsection.res_ubc103}, a search for outlying members of UBC~103 finds a strong concentration of objects with compatible astrometric parameters in the vicinity of the nominal position for NGC~6683. This is a poorly studied cluster, first characterised by \citet{yilmaz66}, who obtained $RGU$ photometry from photographic plates, finding a distance modulus $DM=10.5$ and an estimated earlier spectral type around b4. This is the only dedicated study that we find in the literature. \citet{kharchenko13} estimate a distance $d=1.4$~kpc and an age of only 5.6~Ma. By fitting isochrones to 2MASS photometry, \citet{chen15} derived a much longer $DM \approx 11.6$ ($d\approx2.1$~kpc) and an age $\sim 160\:$Ma.

To characterise the cluster, we performed a Bayesian analysis of the field by using the Virtual Observatory tool Clusterix 2.0 \citep{balaguer20}, an interactive web-based application that calculates the grouping probability of a list of objects by using proper motions and the non-parametric method proposed by \citet{cabreracano90} and described in \citet{galadi98}. We took EDR3 objects with $RUWE<1.4$ and errors in proper motion below $0.5\:$mas$\:\mathrm{a}^{-1}$, with magnitudes in the $10\leq G\leq18$ range. The cluster central concentration seems to lie about 1~arcmin south of the nominal position in SIMBAD, close to the location considered by \citet{yilmaz66}, whose stars 1 and 3-6 all seem to be members. We therefore centred the circles used by Clusterix on RA = 18:42:15, Dec:$-06$:13:40, taking a radius of 10 arcmin for the whole sample and an initial cluster radius of $4\farcm8$.  

\begin{figure}
 \centering
  \resizebox{\columnwidth}{!}{\includegraphics[clip]{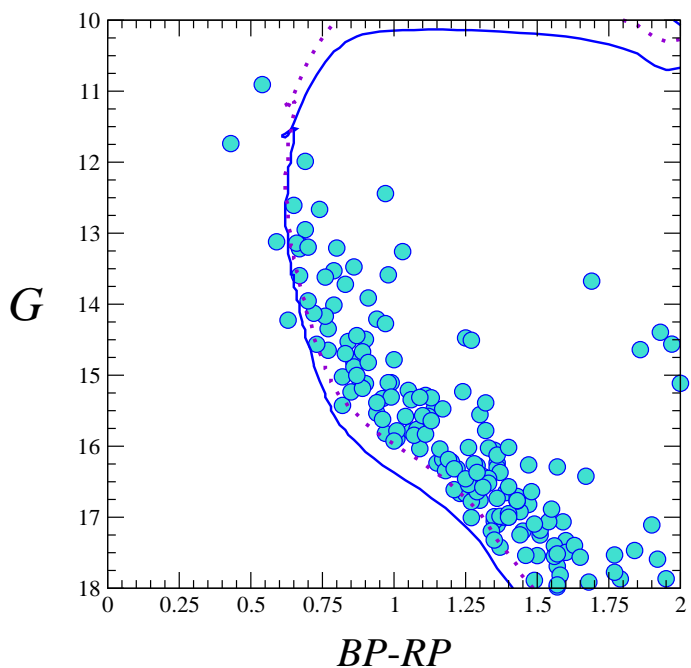}}
 \caption{\textit{Gaia} EDR3 CMD for stars selected as members of NGC~6883. The solid (blue) line is the same Padova isochrone that gives the best fit to UBC~103, i.e. 70~Ma, displaced to $DM=12.4$, but affected by $A_{V}=1.8$. The dotted (purple) line is the same isochrone displaced to $DM=12.0$.}
  \label{N6683_fit}%
 \end{figure}

 Based on an empirical determination of the frequency functions in the vector point diagram \citep{sanders71}, Clusterix assigns each object a probability of belonging to a distinct population. The cluster is clearly noticeable in the vector point diagram and stands out very strongly in the probability distribution. We made a cut in probability that leaves out most objects not related to the cluster and then calculated weighted averages for parallax, $\mu_{\alpha}$ and $\mu_{\delta}$. Finally, we performed a sigma-clipping iterative procedure to clean the sample of outliers. For this, we calculated the weighted average (weighted by their errors) of all astrometric parameters and iteratively removed objects more than $2\:\sigma$ away until convergence was reached.
 
 For all three astrometric parameters, median values are indistinguishable from weighted averages, confirming that this is a very well defined population. The average values are $\mu_{\alpha}=-0.34\pm0.07\:\mathrm{mas}\,\mathrm{a}^{-1}$, $\mu_{\delta}=-2.33\pm0.08\:\mathrm{mas}\,\mathrm{a}^{-1}$ and $\varpi=-0.32\pm0.05\:\mathrm{mas}\,\mathrm{a}^{-1}$ (after zero point correction). These values are all less than one $\sigma$ away from the average values for UBC~103, with the parallax being indistinguishable. UBC~103 and NGC~6683 are therefore twin clusters. Lacking any evolved star, it is difficult to fix the age of NGC~6683, but it is certainly not very different from that of UBC~103. Figure~\ref{N6683_fit} shows the \textit{Gaia} photometry together with the same isochrone that fits UBC~103, i.e. a solar metallicity 70~Ma isochrone. The extinction required is slightly higher, $A_{V}=1.8$. As in the case of UBC~103, the fit is not bad, but a shorter distance modulus $DM = 12.0$ would be favoured from the photometry, if we lacked the \textit{Gaia} parallax. 
 
 In any event, based on \textit{Gaia} data, NGC~6683 seems significantly more distant and younger than assumed by previous works. The membership of CK~Sct in NGC~6683, proposed by \citet{chen15}, is not confirmed by \textit{Gaia} astrometry, as this object has a parallax $\varpi =0.55\pm0.08$ in DR2 and $\varpi =0.45\pm0.02$ in EDR3, indicating a shorter distance, and very different proper motions.

\end{appendix}
\end{document}